\documentclass[twocolumn]{aa} 
\usepackage{longtable}
\usepackage{graphicx}
\usepackage{natbib}
\usepackage{deluxetable}
\usepackage{url}
\usepackage{txfonts}
\begin{document}
   \title{A deep look into the cores of young clusters\thanks{Based on observations made at the ESO La Silla and Paranal Observatory under programmes 67.C-0042, 074.C-0084, and 074.C-0628}}
\subtitle{I. $\sigma-$Orionis}

   \author{H. Bouy\inst{1,2} \thanks{Marie Curie Outgoing International Fellow MOIF-CT-2005-8389}
          \and N. Hu\'elamo\inst{3}
	  \and E.~L. Mart\'\i n\inst{1,4}
          \and F. Marchis\inst{2}
 	  \and D. Barrado y Navascu\'es\inst{3}
          \and J. Kolb\inst{5}
          \and E. Marchetti\inst{5}
          \and M.~G. Petr-Gotzens\inst{5}
          \and M. Sterzik\inst{6}
          \and V.~D. Ivanov\inst{6}
          \and R. K\"ohler\inst{7}
          \and D. N\"urnberger\inst{6}
          }

   \offprints{H. Bouy}

   \institute{Instituto de Astrof\'\i sica de Canarias, C/ V\'\i a L\'actea s/n, E-38205 - La Laguna, Tenerife, Spain\\
     \email{bouy@iac.es}
     \and
     Astronomy Department, University of California, Berkeley, CA 94720, USA\\
     \and
     Laboratorio de Astrof\'\i sica Espacial y F\'\i sica Fundamental (LAEFF-INTA), PO BOX 78, E-28691, Villanueva de la Ca\~nada, Madrid, Spain\\
     \and
     University of Central Florida, Department of Physics, P.O. Box 162385, Orlando, FL 32816-2385, USA\\
     \and
     European Southern Observatory, Karl Schwartzschild Str. 2, D-85748 Garching bei M\"unchen, Germany\\
     \and
     European Southern Observatory, Alonso de Cordova 3107, Vitacura, Casilla 19001, Santiago 19, Chile \\
     \and
     ZAH, Landessternwarte, K\"onigstuhl, D-69117 Heidelberg, Germany
   }

   \date{Received ; accepted }

 
  \abstract
   {Nearby young clusters are informative places to study star formation history. Over the last decade, the $\sigma-$Orionis cluster has been a prime location for the study of young very low mass stars, substellar and isolated planetary mass objects and the determination of the initial mass function.}
   {To extend previous studies of this association to its core, we searched for ultracool members and new multiple systems within the 1\farcm5$\times$1\farcm5 central region of the cluster. }
   {We obtained deep multi-conjugate adaptive optics (MCAO) images of the core of the $\sigma-$Orionis cluster with the prototype MCAO facility MAD at the VLT using the H and K$_{\rm s}$ filters. These images allow us to detect companions fainter by $\Delta$H$\approx$5~mag as close as 0\farcs2 on a typical source with H=14.5~mag. These images were complemented by archival SofI Ks-band images and \emph{Spitzer} IRAC and MIPS mid-infrared images}
   {We report the detection of 2 new visual multiple systems, one being a candidate binary proplyd and the other one a low mass companion to the massive star $\sigma$~Ori~E. Of the 36 sources detected in the images, 25 have a H-band luminosity lower than the expected planetary mass limit for members, and H-K$_{\rm s}$ color consistent with the latest theoretical isochrones. Nine objects have additional \emph{Spitzer} photometry and spectral energy distribution consistent with them being cluster members. One of them has a spectral energy distribution from H to 3.6~$\mu$m consistent with that of a 5.5~M$\rm_{Jup}$ cluster member. Complementary NTT/SofI and \emph{Spitzer} photometry allow us to confirm the nature and membership of two L-dwarf planetary mass candidates.}
   {}


   \maketitle
%

\section{Introduction}
Over the last decade, the $\sigma-$Orionis cluster has become one of the prime locations for the study of brown dwarfs (BDs) and planetary-mass objects (PMOs). It is young (2--3~Myr), free of extinction (A$_{\rm V}<$1~mag) and it has a large low-mass population \citep{1999ApJ...521..671B,2001ApJ...556..830B,2003A&A...404..171B,2004Ap&SS.292..339B,2004AJ....128.2316S,2005MNRAS.356...89K,2007AN....328..917C,2007A&A...470..903C,2008MNRAS.383..375C,2006A&A...460..799G, 2008arXiv0805.2914S}. The {\it Hipparcos} parallax to the central OB pair $\sigma$~Ori~AB is 320$^{+120}_{-90}$~pc, and most previous studies used 350~pc to $\sigma$~Ori~AB as the cluster distance. Other authors estimated  distances at about 390~pc \citep{2008MNRAS.383..750C,2008MNRAS.386..261M}. \citet{2008AJ....135.1616S} recently refined the measurement using main-sequence fitting and derived an improved value of 440~pc for a solar metallicity \citep{2006PhDT........Caballero}. The cluster mass function rises steadily from the very low mass stars through the BDs and into the PMO domain \citep{2001ApJ...556..830B,2007A&A...470..903C}. No obvious discontinuities are seen either at the mass boundary between very low-mass stars and BDs or at the frontier between BDs and PMOs, which is not surprising as the initial mass function does not reflect the onset of nuclear reactions that may have occured during the subsequent evolution for stars and BDs. About two dozens PMOs candidates have been identified in the cluster \citep{2002ApJ...578..536Z,2000Sci...290..103Z,2007A&A...470..903C}. About half of those have been confirmed spectroscopically \citep{2001ApJ...558L.117M,2001A&A...377L...9B,2003ApJ...593L.113M,2000Sci...290..103Z}. 

Evidence of disks has been found around BD and PMO members by the detection of large H$\alpha$ emission, flux excesses at mid-infrared (mid-IR) wavelengths and large-amplitude photometric variability \citep{2003AJ....126.2997B, 2003A&A...404..171B,2003ApJ...592..266M,2006A&A...445..143C, 2004A&A...419..249S, 2008ApJ...672L..49S,2007ApJ...662.1067H}. In particular the very strong H$\alpha$ emission brown dwarf SOri~71, located just above the cluster deuterium burning limit, displays excess flux in IRAC band 4.5 at 8.0~$\mu$m \citep{2007A&A...470..903C}. If the cluster PMOs form in a similar way to the very low-mass stars and BDs, we expect that they would have dusty disks, probably of lower mass scaling with primary mass. \citet{2007ApJ...662.1067H} suggested an increase of the disk frequency  towards low masses in the cluster, with a peak of 40\% in the mass  interval 0.2--0.1~M$_{\sun}$. According to \citet{2007A&A...470..903C}, the disk  rate in the BD domain could be as high as 50\%. \citet{2007A&A...472L...9Z} and \citet{2008ApJ...672L..49S} have reported mid-IR excesses in seven cluster PMOs, indicating that more than  30\% of them have dusty disks. \citet{2008A&A...481..423G} reported that the mass accretion rate of $\sigma-$Orionis members harboring disks is significantly lower than that of e.g $\rho-$Oph members. On the other hand, \citet{2008arXiv0805.2914S} report a significantly larger fraction of accretors than in the neighboring $\lambda-$Orionis association where star formation might have been triggered by a supernova explosion \citep{2001AJ....121.2124D}.

The multiplicity of $\sigma-$Orionis members has been the subject of a number of studies. Using a Fraunhofer micrometer, \citet{1837AN.....14..249S} resolved $\sigma$~Ori~AB for the first time as a close binary (0\farcs26). A few decades later, \citet{1893AN....131..329B} used the micrometer mounted on the 36 inch telescope at the Lick Observatory and confirmed the multiplicity of $\sigma$~Ori~AB. \citet{1904ApJ....19..151F}  identified a possible spectroscopic component in the $\sigma$~Ori~AB binary system. More recently \citet{2005AN....326.1007C} obtained adaptive optics (AO) images of the central region of the cluster, resolving a number of sources. \citet{2005MNRAS.356...89K}  used high resolution spectroscopic measurements of a sample of candidate very low-mass stars and BDs of the association to confirm the youth and membership and to search for spectroscopic binaries. Their preliminary conclusions, based on small-number statistics, shows that the binary fraction among $\sigma-$Orionis very low mass members is higher than that reported for their older field counterparts. Our understanding of the $\sigma-$Orionis cluster was recently complicated by the discovery by \citet{2006MNRAS.371L...6J} and \citet{2007A&A...462L..23S} of two distinct kinematics populations with different ages. However, most of the objects belonging to the older kinematic group are located northward of the central region of the $\sigma-$Orionis cluster, outside the area covered by the present study.

To extend the previous studies of $\sigma-$Orionis to its core we conducted an AO assisted imaging survey of the central part of the $\sigma-$Orionis cluster with the ESO multi-conjugated AO prototype instrument \emph{Multi-Conjugate Adaptive Optics Demonstrator} \citep[hereafter MAD][]{2006SPIE.6272E..21M}. These deep images have a resolution of $\approx$0\farcs1 on a field of view of 1\farcm5$\times$1\farcm5.
 
\section{MCAO Observations}

\subsection{MAD: a multi-conjugate adaptive optics facility at the VLT}
MAD is a prototype instrument performing wide field-of-view, real-time correction for atmospheric turbulence \citep{2006SPIE.6272E..21M}. MAD was built by the European Southern Observatory (ESO) with the contribution of two external consortia to prove the feasibility of MCAO on the sky in the framework of the 2$^{\rm nd}$ generation VLT instrumentation and of the European Extremely Large Telescope \citep[ELT, ][]{2007Msngr.127...11G}. Originally designed as a laboratory experiment, MAD was offered to the community for science demonstration in November 2007 and January 2008 and was installed at the Visitor Focus of the VLT telescope UT3 Melipal. An overview of its performance on the sky is given in \citet{2007Msngr.129....8M, 2008A&A...477..681B}. Its CAMCAO near-IR camera is based on a 2048$\times$2048 pixel HAWAII-2 infrared detector with a pixel scale of 0\farcs028 for a total field of view of 57\farcs3$\times$57\farcs3. At the time of the science demonstration observations, the CAMCAO camera suffered from a form of light leak. The amount of light and the pattern seen on the frames depend on the position of the camera, the DIT and the observing conditions, so that it is not possible to perfectly correct for it. The three wavefront sensors can close the loop on stars brighter than V$\approx$12~mag within a circular area of 2\arcmin\, diameter.

\subsection{Observations}
\begin{figure}
   \centering
   \includegraphics[width=0.45\textwidth]{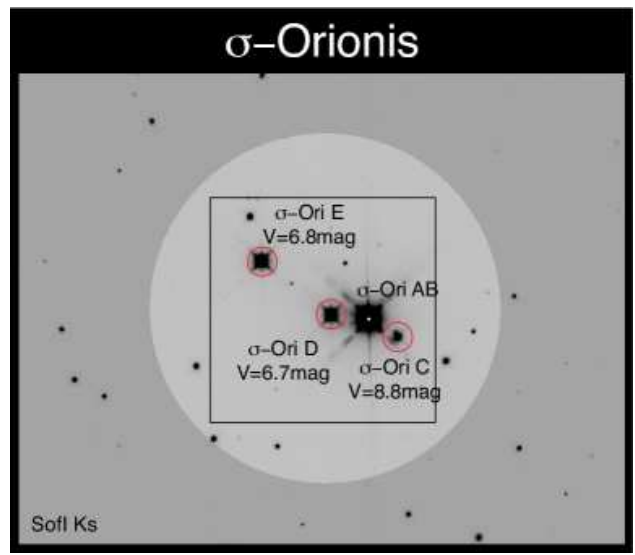}
      \caption{NTT/SOFI K$_{\rm s}$-band mosaic image of the observed field. The field of view within which the wavefront sensing stars for MAD can be selected is represented with a circle of 2\arcmin\, diameter. The 3 wavefront sensing reference stars are indicated with red circles and their names and V-band luminosities overplotted. The 1\farcm5$\times$1\farcm5\, field-of-view of the final images is also represented. North is up and east is left.}
         \label{sofiimage}
   \end{figure}

During the on-sky demonstration run of MAD held in November 2007, a region of ~1\farcm5$\times$1\farcm5\, centered on the $\sigma-$Orionis cluster was observed in the H and K$_{\rm s}$ filters. Figure~\ref{sofiimage} gives an overview of the pointings and of the guide stars used for wavefront sensing. The geometrical distribution of the guide stars is quite asymmetric, leading to non-optimal corrections. 

A set of \verb|NINT|=30 images was obtained by dithering within a box of 15\arcsec\, using the scanning capability of the infrared camera and keeping the adaptive-optics loop closed during the whole operation. The ambient conditions\footnote{At the zenith and in the visible.} during the observations reported by the ESO Ambient Conditions Database are given in Table \ref{ambient}. The conditions were significantly better during the H-band observations, with a coherence time of $\tau_{0}$=3.2~ms, and an average seeing of 0\farcs63 three times better than during the K$_{\rm s}$ band observations, explaining the better quality and sensitivity of the H band images. The exposure time per individual image was 30$\times$0.8~s (\verb|NDIT|$\times$\verb|DIT|), so that the total exposure time for the final mosaics added up to 12~min in each band.

\begin{figure}
   \centering
   \includegraphics[width=0.5\textwidth]{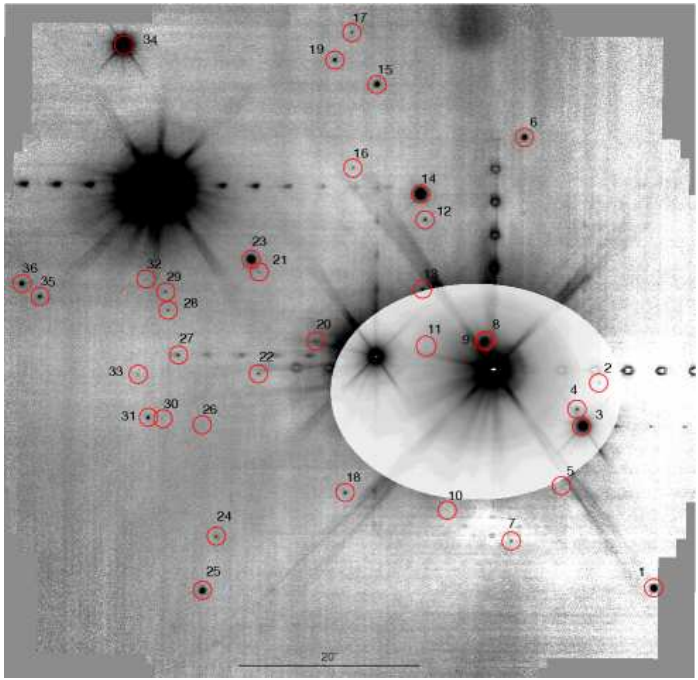}
      \caption{MAD H-band mosaic image with all the detected sources overplotted. The levels have been stretched differently close to the bright OB stars to enhance the contrast. The source numbers are indicated (see Table~\ref{madsources}). North is up and east is left and the scale is indicated.}
         \label{mad_H}
   \end{figure}

The corresponding images were processed with the \emph{Eclipse} reduction package \citep{1997Msngr..87...19D}. They were first dark-subtracted and flat-fielded. The sky contribution was then removed from the input frameset by filtering out low-frequency sky variations from the cube of jittered images. The images were then aligned and stacked to produce the final mosaics. The astrometric solution was computed using isolated and unresolved 2MASS counterparts and is accurate to within 0\farcs1. The final processed mosaics are available upon request from the authors of this article. Figure~\ref{mad_H} shows the final H-band mosaic.

\subsection{MCAO performances}
The Strehl ratio in the K$_{\rm s}$ band ranges from 2\% to 6\%. The undersampling of the PSF in the H band prevents us from computing meaningful Strehl ratios, but the performance are expected to be similar. The quality of the correction closely follows the geometry of the 3 reference stars, and most sources away from the line made by the 3 reference stars are elongated with ellipticities in the range 0.41$\le e\le$0.95. In spite of the low Strehl ratios, the PSFs are much sharper than in the seeing limited SofI images (full width at half maximum FWHM$\approx$0\farcs8, see Section~\ref{sofi}) with an average FWHM of 0\farcs10 in H and 0\farcs15 in K$_{\rm s}$.

\begin{figure}
   \centering
   \includegraphics[width=0.5\textwidth]{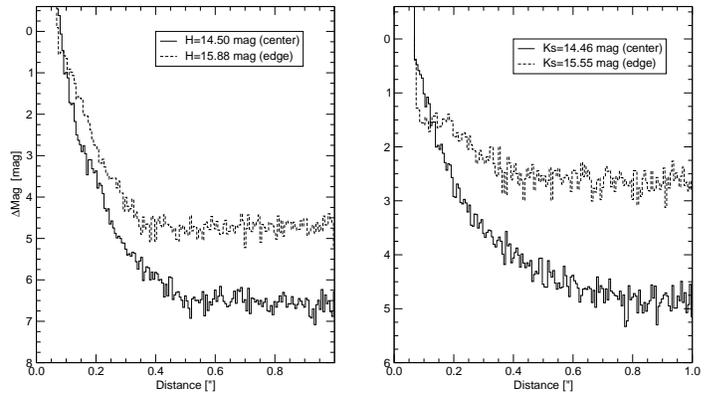}
      \caption{Limit of sensitivity in the H-band (left panel) and K-band (right panel) for 2 different stars located in a region of good AO correction (center) and worse AO correction (edge). The curves have been computed from the 3-$\sigma$ noise of the radial profile of the PSF.}
         \label{limsens}
   \end{figure}

Adaptive-optics provide  not only high spatial resolution but also high-contrast images. To illustrate the performances of the instrument and the limitations of the observations, we computed the limit of sensitivity for two cases: a star with H \& K=14.5~mag located in a region of good AO correction  (hereafter referred to as the ``center'') and a fainter star (H=15.88~mag,K=15.55~mag) located in a region of worse AO correction (hereafter referred to as the ``edge''). The limit of sensitivity was computed using the 3-$\sigma$ standard deviation of the PSF radial profile. Figure \ref{limsens} shows the results. The MAD images allow detection of companions with a magnitude contrast $\Delta m$=4~mag at 0\farcs25 and 0\farcs35 in H and K$_{\rm s}$ respectively on a 14.5~mag star at the center. Sources brighter than $\approx$8~mag were saturated or above the detector linearity limit. We detected sources as faint as K$_{\rm s}$=19.55~mag and H=21.65~mag (3-$\sigma$ detection), therefore well below the deuterium burning limit at the age (1--5~Myr) and distance (440~pc) of the cluster \citep[predicted at H=17.03~mag for the DUSTY models and H=16.87~mag for the COND models, ][]{2000ApJ...542..464C, 2003A&A...402..701B}.

\subsection{Photometry}
Because of the complexity of the MCAO wavefront sensing, the PSF shows spatial variations due to anisoplanetic effects in the AO observations that can affect PSF photometry. To alleviate this problem, we took advantage of the sparsity of the field and performed aperture photometry rather than PSF fitting, except in a few cases of close multiple systems or when the source was located in the halo of a bright neighboring massive star. In these latter cases we extracted the photometry using nearby isolated stars as the reference PSF. The aperture photometry was performed using standard routines with the \emph{daophot} package within IRAF\footnote{IRAF is distributed by the National Optical Astronomy Observatory, which is operated by the Association of Universities for Research in Astronomy (AURA) under cooperative agreement with the National Science Foundation.}, using an aperture of 18~pixels (0\farcs5), and a sky annulus between 20--24~pixels (0\farcs56--0\farcs67). Two well-behaved isolated and unresolved 2MASS sources with clean H-band photometry (quality flag \verb|A|, 2MASS~J05384652-0235479 and 2MASS~J05384746+0235252) were used to derive the photometric instrumental zeropoints given in Table \ref{zp}. The K$_{\rm s}$-band zeropoints were computed using 11 clean and unresolved matches found in the SofI image (see Section \ref{sofi}). Some systematic errors might remain because of the strong anisoplanetism and unaccounted color terms. They are difficult to estimate because of the small number of well behaved counterparts in the 2MASS and SofI catalogs. The relatively small scatter between the MAD K$_{\rm s}$ and SofI K$_{\rm s}$ photometry of the 11 isolated sources in common (well within the uncertainties) suggest that the aperture was chosen large enough that these spatial variations do not affect the final photometry too much. The 36 detected sources and their photometry are reported in Table~\ref{madsources}.

\section{Complementary archival data}
We searched the {\it ESO} and {\it Spitzer} public archives for complementary datasets of the same field. 

\subsection{NTT/SofI images: \label{sofi}}
\begin{figure}
   \centering
   \includegraphics[width=0.5\textwidth]{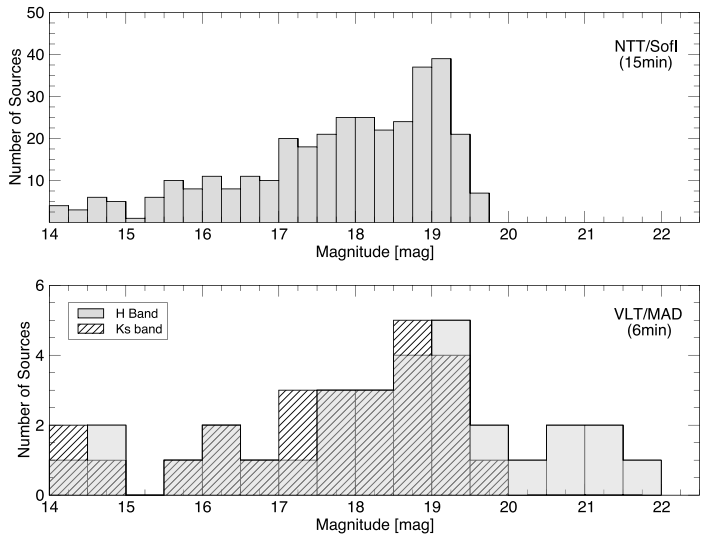}
      \caption{Distribution of magnitude of the sources detected in the SofI (15~min on-source exposure) and MAD (6~min on-source exposure) images. The limits of completeness in the MAD images reach K$_{\rm s}\approx$19.5~mag and H$\approx$20.0~mag. It is not homogeneous over the entire image, as large areas are contaminated by the strong halos of the bright massive stars.\label{limcomp}}
   \end{figure}

The cluster was observed in the K$_{\rm s}$ band with SofI at the NTT on 2001 December 12 (P.I. Testi, Programme 67.C-0042). A set of \verb|NINT|=15 dithered images of 12$\times$5~s (\verb|NDIT|$\times$\verb|DIT|) was obtained that night. We retrieved the data and the corresponding calibration frames and processed them following standard procedures using the recommended \emph{Eclipse} reduction package. The seeing (measured on the image) was 0\farcs8. We extracted the PSF photometry of all the sources brighter than the 3-$\sigma$ noise of the local background using the \citet{2000SPIE.4007..879D} Starfinder code. Using six unresolved 2MASS sources (quality flag \verb|A|) we derived a zeropoint magnitude of 23.96$\pm$0.19~mag. Table~8 is available on-line. The limit of detection of the images is $\approx$20~mag, and the limit of completeness is $\approx$19~mag as illustrated in Fig.~\ref{limcomp}. The detector non-linearity reaches about 3\% at 14000~A.D.U., corresponding to K$_{\rm s}$=13.6~mag.

\subsection{VLT/NACO images:}
\begin{figure}
   \centering
   \includegraphics[width=0.5\textwidth]{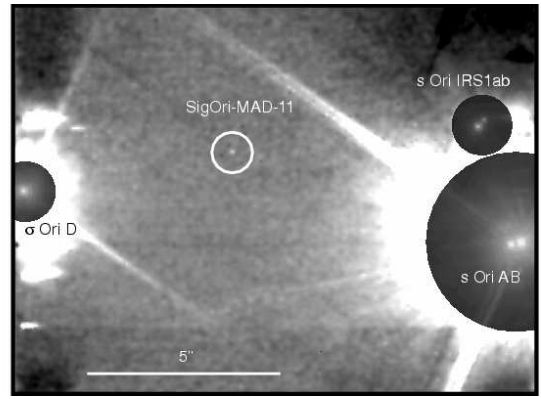}
      \caption{NACO K$_{\rm s}$ image of $\sigma$~Ori~AB, D and IRS1. The image was wavelet filtered to enhance SigOri-MAD-11 detection, indicated with a white circle. East is left and north is up and the scale is shown. The contrast was stretched differently around the bright OB stars to increase the dynamics of the figure. \label{sigoriirs1_naco}}
   \end{figure}

The central stars $\sigma$~Ori~AB and D were observed twice with the AO facility NACO on the VLT (Programs 074.C-0084 and 074.C-0628, P.I. Neuh\"auser). A first set of 4 images of 100$\times$0.345~s (\verb|NDIT|$\times$\verb|DIT|) each was obtained on 2004 October 10 in the K$_{\rm s}$ band with the S13 camera. A second set of 10 dithered images of 10$\times$0.345~s (\verb|NDIT|$\times$\verb|DIT|) was obtained in the narrow band Br$\gamma$ (\verb|NB_2.17|) filter. We retrieved the data and the associated calibration frames and processed them using the recommended \emph{Eclipse} reduction package. The second epoch narrow band images are much shallower than the first ones, and we do not discuss them further. The limited number of images obtained at the first epoch does not allow us to correct perfectly for the bad and hot pixels. The brightness of the massive  OB pair makes it an easy target for NACO and the Strehl ratio was high (60$\sim$65\%, FWHM=0\farcs07). The final image is shown in Fig.~\ref{sigoriirs1_naco} and the astrometric measurements in Table~\ref{astrometry}.

\subsection{{\it Spitzer} data}
\begin{figure}
   \centering
   \includegraphics[width=0.5\textwidth]{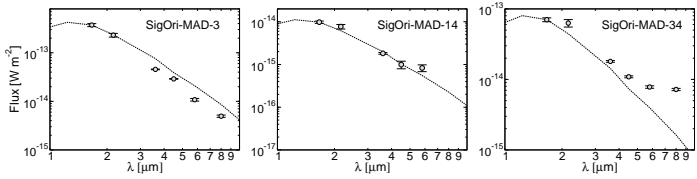}
      \caption{Spectral energy distributions of the MAD stellar sources with {\it Spitzer} IRAC counterparts. The dashed line represents the median SED of confirmed members without excess as derived by \citet{2007ApJ...662.1067H} and normalized to the same H-band flux as the MAD source. SigOri-MAD-34 displays some mid-IR excess indicating the presence of circumstellar material.}
         \label{seds_stellar}
   \end{figure}

The $\sigma-$Orionis cluster was observed with {\it Spitzer} IRAC on 2004 October 08 in the course of program 37 \citep[P.I. Fazio, ][]{2007ApJ...662.1067H} and with IRAC on 2007 April 03 in the course of program 30395 \citep[P.I. Scholz, ][]{2008ApJ...672L..49S}. Program 37 was executed in High Dynamic Range (HDR) mode providing equal numbers of consecutive short and long exposures. Table~\ref{spitzerlog} gives a summary of the observations. We retrieved the calibrated, individual IRAC BCD (Basic Calibrated Data) images and stacked them following the procedures recommended by the Spitzer Science Center (SSC) with the MOPEX software package and the relevant calibration files. The final long exposure mosaics in channel 1 and 3 are made up of images of both programs 37 and 30395 weighted by their exposure times. Program 30395 channel 2 and 4 images do not overlap with our area of interest. The final mosaic images in these 2 bands are therefore made of program 37 images only. The short exposure mosaics allowed us to extract the photometry of bright sources otherwise saturated in the long exposure mosaics. A total of 9 sources detected in the MAD images are also detected in one (or more) {\it Spitzer} IRAC band. We extracted the photometry using standard PSF photometry procedures within the Interactive Data Language. Uncertainties were tentatively estimated from the Poisson noise weighted by the coverage maps of the mosaics, but the presence of the bright and asymmetric halo and ghosts around the massive stars make it difficult to estimate reliable uncertainties. The results are given in Table~\ref{spitzerphot} and Fig.~\ref{seds_stellar}. The cluster was also observed with MIPS on 2004 March 17 in the course of Program 58 (P.I. Rieke). These observations are described in detail in \citet{2007ApJ...662.1067H}. Except for one object (described in Section~\ref{indivtarg}), the coarser resolution and lesser sensitivity of the MIPS images in the vicinity of the massive central OB stars does not allow us to extract any useful MIPS photometry for the MAD sources. 

   \begin{figure}
   \centering
   \includegraphics[width=0.5\textwidth]{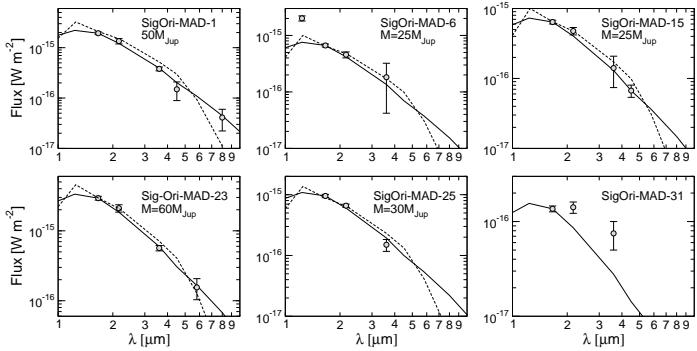}
      \caption{Spectral energy distributions of the MAD substellar candidates with {\it Spitzer} IRAC counterparts. The solid line represent the median SED of confirmed members without excess as derived by \citet{2007ApJ...662.1067H} and normalized to the same H-band flux as the MAD source. The 1~Myr synthetic DUSTY SEDs for the given H-band luminosities and at a distance of 440~pc are overplotted with a dashed line and the corresponding masses indicated. The good match suggests that the sources are indeed members of the association.}
         \label{seds_substellar}
   \end{figure}

\section{Nature of the detections}
In order to rule out the possibility that some of the faint sources detected in the MAD images are artefacts (such as e.g bad pixels, remnants, ghosts, cosmic ray events, etc.), we compare the SofI, NACO and MAD images. Only seven objects detected in the MAD H-band image are not detected in the MAD and SofI K$_{\rm s}$-band images, ruling out the possibility of artefacts for the other 29 objects. 

Two of these seven sources without K$_{\rm s}$ band counterparts in either the MAD or SofI images fall within the field of view of the NACO K$_{\rm s}$ image. Only one is detected (SigOri-MAD-11, see Fig.~\ref{sigoriirs1_naco} and discussion hereafter). The other one (SigOri-MAD-13) falls in an area of the NACO mosaic where only one image was co-added (no overlap with the other 3 images) and where the sensitivity is therefore significantly worse. It is not detected in the SofI image as it falls on a diffraction spike of the bright OB stars where the limit of sensitivity is significantly worse. We estimate the limit of sensitivity at the expected position of SigOri-MAD-13 in both the NACO and SofI images by adding artificial stars of decreasing luminosity until the 3-$\sigma$ detection algorithm misses it. The detection limits measured that way are K$_{\rm s}\approx$17~mag in both the NACO and SofI image, thus just at the limit to detect SigOri-MAD-13. The nature of SigOri-MAD-13 is therefore uncertain and must be confirmed with new images.

The five remaining H-band sources without a MAD, SofI or NACO K$_{\rm s}$ counterpart are either located on diffraction spikes, or in a region where the halos of the bright massive stars are strong, degrading the limit of sensitivity of the SofI or NACO images. These five sources are the faintest of our sample, and with the currently available data we cannot rule out the possibility that they are false positives. However, the presence of confirmed sources with similar magnitudes (e.g SigOri-MAD-26) in both the H and K$_{\rm s}$ images and the quality of the correlation between the PSF of these sources and the PSF of nearby objects (all better than 0.78) suggest that these detections are likely to be real as well. New images with a better signal-to-noise ratio are required to confirm their nature. An upper limit on their Ks band luminosity was derived by adding artificial stars of decreasing luminosity at the expected position of the source until the 3-$\sigma$ detection algorithm no longer detected them.

SigOri-MAD-27 falls close to the position of a saturated bright massive star's remnant spot in both the H and K$_{\rm s}$ MAD images (see Fig.~\ref{mad_H}). The associated H band photometry is therefore unreliable and should be considered with caution. The K$_{\rm s}$ band photometry was measured in the SofI image and is therefore unaffected and reliable.

\section{Multiple systems}

With an average resolution of $\approx$0\farcs1, the MAD images resolve a number of multiple systems. 

\subsection{Previously known multiple systems}

\noindent\emph{$\sigma$~Ori~AB: } The brightest star in $\sigma$--Orionis, which gives its name to the cluster, is a known multiple system made of at least two components, $\sigma$~Ori~A and B \citep{1837AN.....14..249S}. The pair is saturated in all the MAD, NACO K$_{\rm s}$ and SofI images and no useful broad-band photometry can be performed. We used the unsaturated NACO Br$\gamma$ image to measure the relative astrometry of the pair, as reported in Table~\ref{astrometry}.\\

\noindent\emph{$\sigma$~Ori~AD: } \citep{1837AN.....14..249S} The two stars are heavily saturated in the MAD image. $\sigma$~Ori~A is saturated in the NACO image, but not $\sigma$~~Ori~D. The D component is outside the field of view of the unsaturated NACO Br$\gamma$ image. The saturation of the NACO broad-band K$_{\rm s}$ image is not as extensive as in the MAD images and a careful fit of the wings of the saturated PSF of $\sigma$--Ori~A and its Airy rings allow us to measure the relative astrometry of the AD pair.\\

\noindent\emph{$\sigma$~Ori~C:} has been resolved as a wide binary by \citet{2005AN....326.1007C}. It is clearly resolved in the new MAD images. The primary $\sigma$~Ori~Ca is saturated in the MAD H-band but the large separation allows us to accurately measure the photometry of Cb. From the unresolved 2MASS photometry, we derive the relative H-band photometry given in Table~\ref{astrometry}. It is also resolved in the SofI image but the bright Ca primary is saturated. 

\subsection{New multiple systems} 

\begin{figure}
   \centering
   \includegraphics[width=0.5\textwidth]{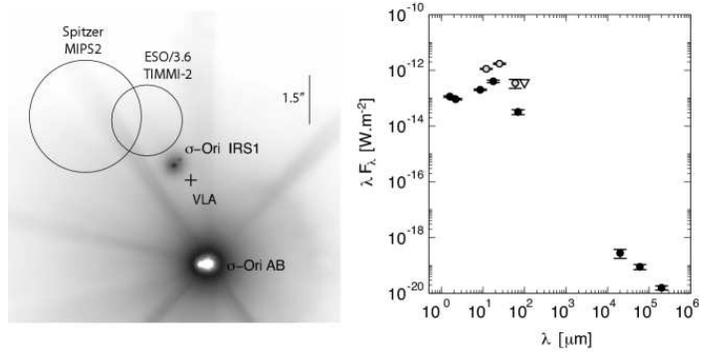}
      \caption{Left panel: MAD H-band image around $\sigma$~Ori~IRS1. The {\it Spitzer} 70~$\mu$m, VLA \citep{1990AJ....100..572D} and TIMMI-2 \citep{2003A&A...405L..33V} detections are overplotted with circles. The size of the circle corresponds to the FWHM of the source in these images, except for the VLA detection which has a much better accuracy ($<$0\farcs1) and is represented with a dimensionless cross. The different positions have been aligned using the MAD H-band image as the reference frame. The uncertainty on the MAD mosaic image astrometric solution is $\approx$0\farcs1. All these detections are closer to $\sigma$~Ori~IRS1 than to $\sigma$~Ori~AB. The scale is indicated. North is up and east is left. Right Panel: Spectral energy distribution of $\sigma$~Ori~IRS1. The IRAS fluxes are represented with grey circles (detections) or a triangle (upper limit). As the source is unresolved in the TIMMI-2, {\it Spitzer} and VLA images, the H and K$_{\rm s}$ fluxes correspond to the combined fluxes of the two components as measured in the MAD and NACO images. }
         \label{sigOriIRS1}
   \end{figure}

\noindent\emph{$\sigma$~Ori~IRS1:} the X-ray, radio, mid- and near-infrared source \citep{2008arXiv0805.0714S,2003A&A...405L..33V,2004A&A...421..715S,2005AN....326.1007C, 1990AJ....100..572D} is resolved in both the MAD and NACO images and we measure its accurate relative astrometry and photometry (see Table \ref{astrometry}). The resolved photometry and colors of the individual component correspond, according to the latest NextGen models of \citet{1998A&A...337..403B}, to masses of 0.47~M$_{\sun}$ and 0.12~M$_{\sun}$, but these values must be considered with caution as previous studies reported a high local extinction toward this source. The object was recently resolved independently by Caballero \& Rebolo (in prep., private communication), and detected with AO in the optical but unresolved by \citet{2008arXiv0805.3162T}. \citet{2003A&A...405L..33V} detected this object in the mid-IR with TIMMI2 at the ESO/3.6m  and associated it with the radio source reported at 2, 6 and 20~cm by \citet{1990AJ....100..572D} and to the mid-IR source IRAS~05362-0237. Based on the mid-IR excess, the displacement between the mid-IR photocenter (associated with a disk) and the radio photocenter (associated with free-free emission from a ionization front) and the presence of processed silicate grains revealed in the mid-IR spectrum, they describe the object as a proplyd, a proto-planetary disk being dispersed by the intense ultraviolet (UV) radiation from $\sigma$~Ori~AB \citep{1996AJ....111..846O}. The source was marginally resolved in their 8.6~$\mu$m images, with a size of $\approx$1\farcs1. \citet{2008arXiv0805.0714S} high resolution Chandra xray observations show that the xray source associated with $\sigma~$Ori~IRS1 is variable and must be a magnetically-active young TTauri star. To further investigate the nature of this source, we retrieved MIPS images of the cluster from the {\it Spitzer} public archive. The observations are described in detail in \citet{2007ApJ...662.1067H}. The central massive pair $\sigma~$Ori~AB is saturated in all IRAC and MIPS 24~$\mu$m images preventing us from extracting useful photometry for $\sigma$~Ori~IRS1.  A bright source is detected in the MIPS 70~$\mu$m image at $\alpha=$05h38min44.9s and  $\delta=$-02\degr35\arcmin56.0\arcsec, i.e only 2\farcs24 from $\sigma$~Ori~IRS1 but 4\farcs9 from $\sigma$~Ori~AB. With a FWHM of $\approx$21\farcs5, it is difficult to associate the mid-IR source with either of these two objects with certainty, but the closer distance to the MAD, TIMMI-2 and VLA sources (see Fig.~\ref{sigOriIRS1}) and the expected much lower 70~$\mu$m flux of the massive pair lead us to associate the MIPS 70~$\mu$m source to $\sigma$~Ori~IRS1. Using PSF photometry, we measure a flux of 752$\pm$150~mJy. This value is not consistent with the IRAS 60~$\mu$m photometry given for the associated source IRAS~05362-0237, which is 6.95~Jy . With a much broader PSF, the IRAS photometry includes the flux of nearby bright nebulosities and is therefore unreliable. Fig.~\ref{sigOriIRS1} shows the relative positions of the MAD, VLA, TIMMI-2 and {\it Spitzer} detections as well as the spectral energy distribution of the source. The formation and evolution of such a system in a Trapezium-like cluster at a projected distance of only $\approx$1200~AU (the physical distance might be much larger) of a pair of massive OB stars make it particularly interesting. One can indeed wonder how such a pair and its circum-binary disk has survived the dynamical interactions commonly occuring in such a cluster for as long as 1$\sim$3~Myr. If confirmed, the nature of this low-mass/very low mass pair will provide  direct proof that relatively wide ($>$100~AU) low mass pairs and their disk can survive the gravitational interactions and photo-evaporation in the early stages of the formation of proto-stellar clusters.\\

\noindent\emph{$\sigma$~Ori~E:} is resolved in the MAD images. The two components are saturated in the H-band image. The secondary is not saturated in the K$_{\rm s}$-band image, but the heavy saturation of the primary and its halo prevent us from making any accurate measurement of the photometry of the secondary. Table~\ref{astrometry} gives approximate measurements of the separation and position angle. We tentatively derive an upper limit on the Ks-band luminosity of the companion by adding an artificial PSF of increasing luminosity at its diametrically opposed position until the luminosity matches that of the companion. The luminosity roughly estimated this way (Ks$\approx$10$\sim$11~mag) corresponds to an estimated mass of 0.4$\sim$0.8~M$\sun$. $\sigma$~Ori~E was suspected to have a low mass companion for several decades. \citet{1974ApJ...191L..95W} had noticed a peculiar variable H$\alpha$ emission in the $\sigma$~Ori~E spectrum and interpreted it (among other hypotheses) as the effect of the possible presence of a very low mass companion. Using \emph{uvby} beta light curves, \citet{1976Natur.262..116H} later suggested that $\sigma$~Ori~E was indeed a binary system, and put an upper limit on the mass of the companion at M$<$0.1~M$_{\sun}$ for inclination $i>$45\degr. The source displayed repeated X-ray flares \citep[reported with {\it ROSAT}, {\it XMM-Newton} and {\it Chandra}, respectively by][]{2004A&A...418..235G,2004A&A...421..715S, 2008arXiv0805.0714S}. The X-ray activity was suspected by several of these authors to be in part due to the presence of an unseen low mass companion. The MAD images show that a low mass companion is indeed present next to the massive helium-strong $\sigma$~Ori~E star and could be responsible for the observed X-ray activity.

\section{Substellar and isolated planetary mass candidates}

\begin{figure}
   \centering
   \includegraphics[width=0.5\textwidth]{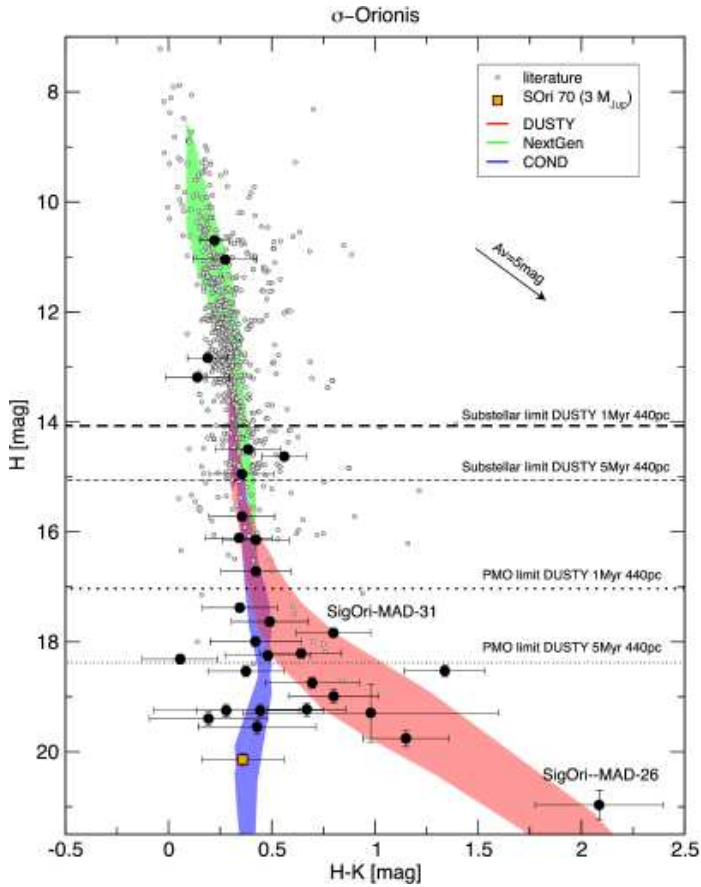}
   \caption{H vs H-K color-magnitude diagram of the $\sigma$-Orionis cluster. The MAD measurements are represented by black circles. The DUSTY, NextGen and COND isochrones between 1--5~Myr and 350--440~pc are represented in red, green and blue shaded areas, respectively. Measurements for cluster members from the literature are overplotted as grey circles \citep{2008A&A...478..667C,2007A&A...470..903C,2004A&A...419..249S,2004Ap&SS.292..339B,2004AJ....128.2316S}. The confirmed PMO SOri~70 \citep{2008A&A...477..895Z} is represented with an orange square. The substellar and deuterium burning limit from the DUSTY models for 1~Myr and 5~Myr at a distance of 440~pc as well as a A$_{\rm V}$=5~mag reddening vector  are  represented.\label{cmd}}
   \end{figure}

Fig. \ref{cmd} shows a color-magnitude diagram of all the sources detected in the MAD images, as well as catalogs of cluster members from the literature for comparison. A number of sources have luminosities and colors consistent with the cluster isochrones and with substellar and planetary masses. With only two bands and no comparison fields away from the cluster it is difficult to assess the level of contamination. Using the model of stellar population synthesis of the Galaxy of \citet{2003A&A...409..523R} we find that the expected contamination by background giants or foreground dwarfs in the field of view of the MAD images must be very low. \citet{2008A&A...488..181C} recently estimated that the number of expected field L and T-dwarf contaminants toward the cluster adds up to $\approx$550 objects per 1deg$^{2}$, corresponding to $<$0.4 contaminants in the case of our study. The contamination by extragalactic sources is expected to be much higher, as illustrated by the recent survey of the cluster by \citet{2007ApJ...662.1067H}. With the current data, it is not possible to estimate the contamination by extragalactic sources among the MAD detections. We nevertheless note that none of the sources identified in the MAD images is extended while most extragalactic sources rejected by \citet{2007ApJ...662.1067H} in their analysis were extended. Additional observations are required to confirm the membership of the new candidates.

Using the {\it Spitzer} photometry, we tentatively assess further the nature of the nine sources with mid-IR counterparts. Figure~\ref{seds_stellar} shows the spectral energy distributions (SED) of the stellar candidate members. All but SigOri-MAD-34 match very well the median SED of cluster members measured by \citet{2007ApJ...662.1067H}, ruling out the possibility of extragalactic contaminants. SigOri-MAD-34 displays a mid-IR excess most likely related to a circumstellar disk and providing further evidence of its youth and membership of the association. Figure~\ref{seds_substellar} shows the SEDs of all the MAD substellar candidates with a {\it Spitzer} counterpart.  All but SigOri-MAD-31 also match very well the median SED of cluster members and the synthetic DUSTY SEDs of the corresponding H-band luminosity at the exact distance of the cluster sequence \citep[440~pc, ][]{2008AJ....135.1616S} and for an age of 1~Myr. These SEDs could be equally well fitted by a foreground late-M field dwarf at a distance of $\approx$100~pc, preventing us from drawing any firm conclusion regarding their membership of the association. Spectroscopy and proper motion measurements are required to confirm their nature.  SigOri-MAD-6 (2MASS J05384454-0235349) was suggested to be a background A-F star or an extragalactic source by \citet{2007AN....328..917C} based on its blue J-K$_{\rm s}$ color. The new 3.6~$\mu$m photometric measurement allows us to rule out the extragalatic contaminant hypothesis. The J-band photometric measurement reported by \citet{2007AN....328..917C} is $\approx$0.7~mag brighter than the expected J-band luminosity of a 0.025~M$\sun$ cluster member. All the other measurements (H, K$_{\rm s}$ and 3.6~$\mu$m) are in good agreement with the luminosities expected for a 0.025~M$\sun$ cluster member as shown in Fig.~\ref{seds_substellar}. The J-band luminosity discrepancy could be due to underestimated errors on the photometric measurement in the proximity of the bright central OB pair (Caballero, private communication). Until new J-band measurements are obtained, SigOri-MAD-6 remains a good brown dwarf member candidate.

\begin{figure}
   \centering
   \includegraphics[width=0.5\textwidth]{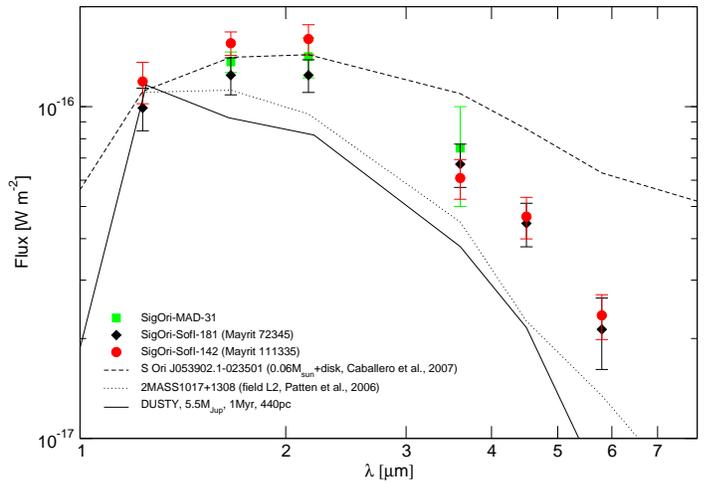}
      \caption{Spectral energy distributions of the substellar candidates with SofI and {\it Spitzer} IRAC counterparts (SigOri-SofI-181, black diamonds, and SigOri-SofI-142, red dots), as well as SigOri-MAD-31 (green squares). The line represents the 1~Myr DUSTY SED of a $\approx$5.5~M$_{\rm Jup}$ object at a distance of 440~pc. The dotted line represents the the SED of the field L2 dwarf 2MASS~J1017+1308 \citep{2006ApJ...651..502P} normalized to the average J-band fluxes of the two candidates. The dashed line represents the SED of the 0.060 brown dwarf with a disk S~Ori~J053902.1-023501 as reported by \citet{2007A&A...470..903C} and normalized to the average J-band fluxes of the two candidates. The three L-dwarf candidates display a clear mid-IR excess when compared to the field L2 dwarf or the DUSTY photospheric model. Their SED look like lower luminosity analogues of the substellar member S~Ori~J053902.1-023501  \citep{2007A&A...470..903C}.}
         \label{seds_sofi}
   \end{figure}

Mayrit~72345 and Mayrit~111335 are two L-dwarf candidate members reported by \citet{2007AN....328..917C} using optical, near-infrared and xray photometry. The two sources are detected in the SofI Ks-band image and in the {\it Spitzer} IRAC1, 2 and 3 channels (respectively SigOri-SofI-181 and SigOri-SofI-142, Table~\ref{spitzersofi} and \ref{sofisources}). The SofI photometry is in good agreement with the \citet{2007AN....328..917C} photometry within the uncertainties. Their SEDs, shown in Fig.~\ref{seds_sofi}, allow us to rule out the possibility that these two sources are extragalactic contaminants. They are inconsistent with a purely photospheric emission and display some excess in the near- and mid-IR most likely related to the presence of circumstellar material, suggesting their youth and membership of the association. In the near-IR, these two objects look like lower luminosity analogues of the substellar member S~Ori~J053902.1-023501 \citep{2007A&A...470..903C}. Their mid-IR excess is less than that reported for S~Ori~J053902.1-023501. \citet{2007AN....328..917C} associated Mayrit~72345 with the X-ray source NX~77 detected with {\it XMM-Newton} by \citet{2006A&A...446..501F}.  Figure~\ref{spitzersofi} shows that their SED is very similar to that of SigOri-MAD-31, adding further evidence that this latter source is likely to be a very low mass substellar member. The J and H-band luminosities of these 3 objects are well matched by a DUSTY SED with a mass of $\approx$5.5~M$\rm_{Jup}$ at a distance of 440~pc and at an age of 1~Myr (Fig.~\ref{spitzersofi} and \ref{seds_substellar}). At 5~Myr, these luminosities correspond to a mass of $\approx$7~M$\rm_{Jup}$, as described in \citet{2007AN....328..917C}. This estimate is only tentative as the objects display some near-IR excess.

\section{Notes on individual targets \label{indivtarg}}

\noindent\emph{SigOri-MAD-11} is detected in the MAD  (12-$\sigma$) and NACO  (3-$\sigma$)  images, confirming that it is a real detection. The two observations are separated by $\approx$3~yr and we tried to measure the proper motion relative to the closest unsaturated object, $\sigma$~Ori~IRS1a. The large uncertainties \citep[dominated by the poorly calibrated camera distortions, of the order of 1\% of the pixel scale for CONICA and unknown for CAMCAO see e.g.][]{2007arXiv0706.2613S, 2007A&A...474..273E} and the small proper motion of the association (only $\approx$6~mas/yr, corresponding to $\approx$18~mas between the two epochs, or 1.35 CONICA pixels and 0.9 CAMCAO pixels) make these measurements inconclusive. New measurements covering a longer timescale will be required to confirmed that this source is co-moving with the nearby massive cluster members.
\vspace{0.5cm}

\noindent\emph{SigOri-MAD-26} is the faintest and reddest object with a detection in both H and K$_{\rm s}$. It has a counterpart in the SofI image. Its luminosity and color are consistent with the DUSTY 1~Myr isochrone at 440~pc within the uncertainties, but are largely inconsistent with the color of the confirmed member S~Ori~70, which lies on the much bluer COND isochrones.  This source is therefore unlikely to be a very low mass substellar member of the association. The K-band peak in the SED could be that of a low redshift (z$\approx$0.5) galaxy. We nevertheless note that an average size galaxy (10\,000~light year diameter) at that distance would have been easily resolved by our MCAO images. The object lies in the direction of the reddening vector, suggesting that it is most likely an extinguished source.
\vspace{0.5cm}

\noindent\emph{SigOri-MAD-34 (Mayrit~53049)} is clearly detected in all IRAC four bands, and displays some excess at long wavelengths indicating the presence of circumstellar material and adding further evidence that the object is young and a member of the association. Assuming an age of 1~Myr and a distance of 440~pc, its H and K$_{\rm s}$-band luminosities correspond to a mass of $\approx$0.45~M$_{\sun}$ according to the NextGen models.
\vspace{0.5cm}

\section{Spatial distribution of BD and PMOs}

Using DENIS, 2MASS and previously published catalogs, \citet{2008MNRAS.383..375C, 2007AN....328..917C} recently concluded that there is an apparent deficit of very low mass stars and high-mass BDs in the central 4\arcmin\, of the cluster. The relatively large number of faint objects detected in our new images of the central part of the cluster suggest that a large fraction of very low mass members might have been missed by the previous survey because of the very bright central massive stars. Their presence affected all the previous seeing limited observations, making the detection of very low-mass cluster members in this region very difficult, if not impossible. The unprecedented dynamic range provided by MAD allows us to detect a number of faint sources that, if confirmed as cluster members, could fill the very low mass end of the central cluster population. Additional observations are required to confirm the membership and nature of the candidates and provide a quantitative answer to that question.

\section{Conclusions and future prospects}
Using multi-conjugate AO images of the core of the $\sigma-$Orionis cluster, we have identified 6 new BD candidates and 25 planetary mass candidates. Five of these have additional mid-IR {\it Spitzer} IRAC photometry consistent with that of sub-stellar members. With the current data, it is not possible to conclude on their membership of the association as they could also be foreground late-M dwarfs. The candidate proplyd $\sigma$~Ori~IRS1 is resolved as a binary. The high spatial resolution MAD images resolve 5 pairs, including 2 previously unknown ones. The results presented in this paper illustrate the capacity of multi-conjugate AO to probe the immediate vicinity of young massive OB stars in great detail. Using complementary archival SofI and {\it Spitzer} images, we were able to confirm the membership of two L-dwarf candidate members that exhibit near- and mid-IR excess associated with a disk. One planetary mass candidate newly detected in the MAD images (SigOri-MAD-31) displays a SED very similar to these latter two, suggesting that it is also a L-dwarf member of the association harboring a disk. The presence or absence of very low mass stars, BDs and planetary mass objects close to massive stars provides novel constraints on the models of formation. Follow-up observations of the candidates are very much needed to confirm the nature and membership of the new candidates, and to provide quantitative feedback on the models of formation.

\begin{acknowledgements}
The authors are grateful to Paola Amico for her support at ESO. We thank J.~A. Caballero and V. B\'ejar for fruitful discussions, suggestions and comments on the manuscript. We thank our anonymous referee for her/his review of the manuscript. H. Bouy acknowledges the funding from the European Commission's Sixth Framework Program as a Marie Curie Outgoing International Fellow (MOIF-CT-2005-8389). F. Marchis work was supported by the National Science Fundation Science and Technology Center for Adaptive Optics, and managed by the University of California at Santa Cruz under cooperative agreement No AST-9876783. Nuria Hu\'elamo and David Barrado y Navascu\'es are funded by Spanish grants MEC/ESP2007-65475-C02-02, MEC/Consolider-CSD2006-0070 and CAM/PRICIT-S-0505/ESP/0361. This work is based on observations obtained with the MCAO Demonstrator (MAD) at the VLT (ESO Public data release), which is operated by the European Southern Observatory. The MAD project is led and developed by ESO with the collaboration of the INAF-Osservatorio Astronomico di Padova (INAF-OAPD) and the Faculdade de Ci\^encias de Universidade de Lisboa (FCUL). Based on observations made with ESO Telescopes at the La Silla or Paranal Observatories under programmes 67.C-0042, 074.C-0084, and 074.C-0628. This publication makes use of data products from the Two Micron All Sky Survey, which is a joint project of the University of Massachusetts and the Infrared Processing and Analysis Center/California Institute of Technology, funded by the National Aeronautics and Space Administration and the National Science Foundation. This work has made use of the Vizier Service provided by the Centre de Donn\'ees Astronomiques de Strasbourg, France \citep{Vizier}. This research used the facilities of the Canadian Astronomy Data Centre operated by the National Research Council of Canada with the support of the Canadian Space Agency. 

\end{acknowledgements}

\bibliographystyle{aa}
\bibliography{mybiblio}


\begin{center}
\begin{table}
\caption{Ambient conditions at the zenith and in the visible during the observations}
\label{ambient}
\begin{tabular}{lcccc}\hline\hline
Filter        & date                 & airmass       & seeing     & $\tau_{0}$      \\
              & [UT]                 &               & [\arcsec]  & [ms]           \\
\hline
K$_{\rm s}$            & 2007-11-25 06:33 & 1.09     & 1\farcs90$\pm$0\farcs13 & 0.9$\pm$0.1  \\
H             & 2007-12-01 07:37 & 1.23     & 0\farcs63$\pm$0\farcs05 & 3.2$\pm$0.3  \\
\hline
\end{tabular}
\end{table}
\end{center}

\begin{center}
\begin{table}
\caption{Instrumental zeropoints for the MAD observations}
\label{zp}
\begin{tabular}{lll}\hline\hline
Method    &   Filter    & Zeropoint [mag]   \\
\hline
Aperture  & H         & 25.25$\pm$0.07 \\
Aperture  & K$_{\rm s}$         & 24.62$\pm$0.10 \\
PSF       & H         & 25.30$\pm$0.07 \\
PSF       & K$_{\rm s}$         & 24.56$\pm$0.10 \\
\hline
\end{tabular}
\end{table}
\end{center}

\begin{table*}
\centering
\caption{Catalog of sources detected in the MAD images \label{madsources}}
\begin{tabular}{lccrrcc}\hline\hline
Name        & RA (J2000)  & Dec (J2000) & H [mag] & K$_{\rm s}$ [mag] & Other name & Comment \\
\hline
{\it SigOri-MAD-1} & 05:38:43.5 & -02:36:23.7 & 14.95$\pm$0.07 & 14.59$\pm$0.14 &      & Spitzer, 0.050~M$\sun$     \\
{\bf SigOri-MAD-2} & 05:38:43.9 & -02:36:01.5 & 17.99$\pm$0.09 & 17.57$\pm$0.20 &      &      \\
SigOri-MAD-3 & 05:38:44.1 & -02:36:06.3 & \nodata & 9.01$\pm$0.10 & $\sigma$~Ori~C, Mayrit~11238 & Spitzer     \\
{\it SigOri-MAD-4} & 05:38:44.1 & -02:36:04.4 & 14.63$\pm$0.07 & 14.07$\pm$0.10 &      &      \\
{\bf SigOri-MAD-5} & 05:38:44.2 & -02:36:12.7 & 20.81$\pm$0.07 & $<$19.0 &      & No SofI     \\
{\it SigOri-MAD-6} & 05:38:44.5 & -02:35:35.0 & 16.11$\pm$0.07 & 15.77$\pm$0.14 & 2MASS~J05384454-0235349 & Spitzer, 0.025~M$\sun$     \\
{\bf SigOri-MAD-7} & 05:38:44.6 & -02:36:18.7 & 18.99$\pm$0.12 & 18.19$\pm$0.18 &      &      \\
SigOri-MAD-8 & 05:38:44.8 & -02:35:56.9 & 12.84$\pm$0.07 & 12.65$\pm$0.07 & $\sigma$~Ori~IRS1~B     &  NACO    \\
SigOri-MAD-9 & 05:38:44.8 & -02:35:57.1 & 10.70$\pm$0.07 & 10.48$\pm$0.01 & $\sigma$~Ori~IRS1~A       & NACO     \\
{\bf SigOri-MAD-10} & 05:38:45.0 & -02:36:15.4 & 21.65$\pm$0.57 & $<$19.0 &      & No SofI     \\
{\bf SigOri-MAD-11} & 05:38:45.2 & -02:35:57.7 & 19.29$\pm$0.53 & 18.31$\pm$0.32 &      & No SofI/ NACO     \\
{\bf SigOri-MAD-12} & 05:38:45.2 & -02:35:44.0 & 18.21$\pm$0.08 & 17.57$\pm$0.17 &      &      \\
{\bf SigOri-MAD-13} & 05:38:45.2 & -02:35:51.5 & 18.54$\pm$0.10 & 17.19$\pm$0.17 &      & No SofI/No NACO     \\
SigOri-MAD-14 & 05:38:45.2 & -02:35:41.2 & 13.19$\pm$0.07 & 12.70$\pm$0.14 & Mayrit~21023 & Spitzer      \\
{\it SigOri-MAD-15} & 05:38:45.6 & -02:35:29.3 & 16.15$\pm$0.07 & 15.72$\pm$0.14 &      & Spitzer, 0.025~M$\sun$     \\
{\bf SigOri-MAD-16} & 05:38:45.7 & -02:35:38.4 & 19.39$\pm$0.12 & 19.20$\pm$0.26 &      &      \\
{\bf SigOri-MAD-17} & 05:38:45.7 & -02:35:23.7 & 19.23$\pm$0.12 & 18.56$\pm$0.15 &      &      \\
{\bf SigOri-MAD-18} & 05:38:45.8 & -02:36:13.5 & 18.26$\pm$0.09 & 17.77$\pm$0.19 &      &      \\
{\bf SigOri-MAD-19} & 05:38:45.9 & -02:35:26.7 & 17.38$\pm$0.08 & 17.03$\pm$0.16 &      &      \\
{\bf SigOri-MAD-20} & 05:38:46.0 & -02:35:57.1 & 18.53$\pm$0.10 & 18.16$\pm$0.16 &      &      \\
{\bf SigOri-MAD-21} & 05:38:46.4 & -02:35:49.7 & 20.07$\pm$0.17 & $<$19.0 &      & No SofI     \\
{\bf SigOri-MAD-22} & 05:38:46.4 & -02:36:00.7 & 19.25$\pm$0.12 & 18.80$\pm$0.28 &      &      \\
{\it SigOri-MAD-23} & 05:38:46.5 & -02:35:48.3 & 14.50$\pm$0.07 & 14.11$\pm$0.14 & 2MASS~J05384652-0235479 & Spitzer, 0.060~M$\sun$     \\
{\bf SigOri-MAD-24} & 05:38:46.7 & -02:36:18.3 & 18.75$\pm$0.10 & 18.05$\pm$0.21 &      &      \\
{\it SigOri-MAD-25} & 05:38:46.8 & -02:36:24.1 & 15.72$\pm$0.07 & 15.37$\pm$0.14 &      & Spitzer, 0.030~M$\sun$     \\
{\bf SigOri-MAD-26} & 05:38:46.8 & -02:36:06.2 & 20.97$\pm$0.27 & 18.88$\pm$0.15 &      &      \\
{\bf SigOri-MAD-27} & 05:38:47.0 & -02:35:58.7 & 18.32$\pm$0.09 & 18.26$\pm$0.16 &      & H-band uncertain  \\
{\bf SigOri-MAD-28} & 05:38:47.1 & -02:35:53.8 & 19.25$\pm$0.11 & 18.97$\pm$0.33 &      &      \\
{\bf SigOri-MAD-29} & 05:38:47.1 & -02:35:51.9 & 19.55$\pm$0.13 & 19.12$\pm$0.25 &      &      \\
{\bf SigOri-MAD-30} & 05:38:47.1 & -02:36:05.6 & 21.39$\pm$0.44 & $<$19.0 &      & No SofI     \\
{\bf SigOri-MAD-31} & 05:38:47.2 & -02:36:05.4 & 17.83$\pm$0.08 & 17.04$\pm$0.16 &      & Spitzer, 7~M$\rm_{Jup}$     \\
{\bf SigOri-MAD-32} & 05:38:47.2 & -02:35:50.5 & 21.01$\pm$0.07 & $<$19.5 &      & No SofI      \\
{\bf SigOri-MAD-33} & 05:38:47.3 & -02:36:00.8 & 19.75$\pm$0.15 & 18.60$\pm$0.15 &      &      \\
SigOri-MAD-34 & 05:38:47.4 & -02:35:25.2 & 11.05$\pm$0.07 & 10.42$\pm$0.14 & Mayrit~53049     &  Spitzer    \\
{\bf SigOri-MAD-35} & 05:38:48.0 & -02:35:52.4 & 17.63$\pm$0.08 & 17.14$\pm$0.17 &      &      \\
{\bf SigOri-MAD-36} & 05:38:48.1 & -02:35:51.0 & 16.72$\pm$0.07 & 16.30$\pm$0.15 &      &      \\
\hline
\end{tabular}

Note -- Objects with H-band luminosity corresponding to a PMO (defined here as M$\le$0.012~M$_{\sun}$) are indicated in bold face. Objects with H-band luminosity corresponding to a BD (defined here as 0.012$<$M$\le$0.072~M$_{\sun}$) are indicated in italic. The errors include the zeropoint uncertainties.
\end{table*}

\begin{center}
\begin{table*}
\caption{Log of Spitzer observations}
\label{spitzerlog}
\begin{tabular}{lcccc}\hline\hline
Program ID  &   P.I.  & Date Obs.    & Exposure time & Number of frames  \\
            &         & [DD-MM-YYYY] & [s]           & \\
\hline
37          & Fazio   & 08-10-2004 & 1.2/30    & 270/270 \\
30395       & Scholz  & 03-04-2007 & 100         &  24 \\
\hline
\end{tabular}
\end{table*}
\end{center}

\begin{table*}
\caption{Spitzer IRAC photometry of sources detected in the MAD images \label{spitzerphot}}
\begin{tabular}{lccrrc}\hline\hline
Name           & 3.6~$\mu$m    & 4.5~$\mu$m    & 5.8~$\mu$m    & 8.0~$\mu$m    \\
               & [mJy]         & [mJy]         & [mJy]         & [mJy]         \\
\hline
SigOri-MAD-1   & 0.46$\pm$0.04 & 0.22$\pm$0.09 & 0.10$\pm$0.04 & 0.11$\pm$0.05    \\
SigOri-MAD-3   & 54.6$\pm$0.04 & 43.0$\pm$0.04 & 20.9$\pm$0.5  & 13.19$\pm$0.7     \\
SigOri-MAD-6   & 0.22$\pm$0.17 & \nodata      & \nodata       & \nodata           \\
SigOri-MAD-14  & 2.19$\pm$0.10 & 1.49$\pm$0.30 & 1.60$\pm$0.30 & \nodata           \\
SigOri-MAD-15  & 0.17$\pm$0.08 & 0.10$\pm$0.02 & \nodata       & \nodata           \\
SigOri-MAD-23  & 0.68$\pm$0.06 & \nodata       & 0.3$\pm$0.1   & \nodata           \\
SigOri-MAD-25  & 0.18$\pm$0.04 & 0.27$\pm$0.05  & \nodata      & \nodata           \\
SigOri-MAD-31  & 0.09$\pm$0.03 & \nodata       & \nodata       & \nodata           \\
SigOri-MAD-34  & 21.2$\pm$0.4  & 16.3$\pm$0.5  & 15.1$\pm$0.6  & 19.1$\pm$0.7      \\
\hline
\end{tabular}
\end{table*}

\begin{table*}
\caption{Spitzer IRAC photometry of the two L-dwarf candidates detected in the SofI images \label{spitzersofi}}
\begin{tabular}{lccrrcc}\hline\hline
Name           & 3.6~$\mu$m    & 4.5~$\mu$m    & 5.8~$\mu$m    & 8.0~$\mu$m  & other   \\
               & [mJy]         & [mJy]         & [mJy]         & [mJy]       & name   \\
\hline
SigOri-SofI-142 & 0.073$\pm$0.010 & 0.070$\pm$0.010 & 0.045$\pm$0.007 & \nodata & Mayrit~111335 \\
SigOri-SofI-181 & 0.080$\pm$0.012 & 0.067$\pm$0.010 & 0.041$\pm$0.010 & \nodata & Mayrit~72345 \\
\hline
\end{tabular}
\end{table*}
\begin{center}
\begin{table*}
\caption{Relative astrometry and photometry of the multiple systems}
\label{astrometry}
\begin{tabular}{lcccccc}\hline\hline
System                 & date  UT             & Separation      & P.A.          & $\Delta$H   &  $\Delta$K$_{\rm s}$ & Instrument \\
                       & [YYYY-MM-DD HH:MM]   & [mas]           & [\degr]       & [mag]         &  [mag]    & \\
\hline
$\sigma$-Ori~AB        & 2004-10-11 09:44     & 255.7$\pm$1.8   & 100.9$\pm$0.4 & \nodata       & \nodata   & NACO    \\
$\sigma$-Ori~AD        & 2004-10-11 09:44     & 13037.2$\pm$27  &  84.1$\pm$0.4 & \nodata       & \nodata   & NACO    \\
$\sigma$-Ori~IRS1~AB   & 2004-10-11 09:44     & 236.3$\pm$2.4   & 318.1$\pm$0.4 & \nodata       & 2.19$\pm$0.01 & NACO    \\
$\sigma$-Ori~IRS1~AB   & 2007-12-01T07:23     & 242.9$\pm$3.6   & 317.0$\pm$0.7 & 2.13$\pm$0.10 & 2.17$\pm$0.07 & MAD     \\
$\sigma$-Ori~Cab       & 2007-12-01T07:23     & 1991.5$\pm$3.9  & 11.5$\pm$0.7  & 5.50$\pm$0.07 & 5.05$\pm$0.14 & MAD    \\
$\sigma$-Ori~Eab       & 2007-12-01T07:23     & $\approx$330    & $\approx$391  & \nodata       & \nodata   & MAD  \\ 
\hline
\end{tabular}

Note -- In addition to the uncertainties on the measurement, these values include errors related to the camera distortions \citep[of the order of 1\% of the pixel scale for NACO, see e.g ][ not calibrated but assumed to be of the same order for MAD]{2007arXiv0706.2613S,2007A&A...474..273E}.
\end{table*}
\end{center}

\begin{center}
\begin{table*}
\caption{NTT/SofI Ks-band photometry}
\label{sofisources}
\begin{tabular}{lccccc}\hline\hline
Name        & RA (J2000)  & Dec (J2000) & Ks [mag] & Nearest 2MASS & Nearest Mayrit \\
\hline
SigOri-SofI-1 & 05:38:33.3 & -02:36:17.0 & 13.05 & 05383335-0236176 &   \\
SigOri-SofI-2 & 05:38:33.3 & -02:36:17.1 & 10.99 & 05383335-0236176 &   \\
SigOri-SofI-3 & 05:38:33.3 & -02:36:44.2 & 18.02 &   &   \\
SigOri-SofI-4 & 05:38:33.5 & -02:34:53.1 & 18.81 &   &   \\
SigOri-SofI-5 & 05:38:33.5 & -02:35:40.6 & 18.17 &   &   \\
SigOri-SofI-6 & 05:38:33.5 & -02:36:05.0 & 18.84 &   &   \\
SigOri-SofI-7 & 05:38:33.6 & -02:35:00.3 & 17.38 &   &   \\
SigOri-SofI-8 & 05:38:33.7 & -02:36:47.5 & 14.71 & 05383376-0236479 &   \\
SigOri-SofI-9 & 05:38:33.7 & -02:36:57.6 & 17.60 &   &   \\
SigOri-SofI-10 & 05:38:33.7 & -02:36:58.3 & 18.88 &   &   \\
SigOri-SofI-11 & 05:38:33.9 & -02:35:31.0 & 18.44 &   &   \\
SigOri-SofI-12 & 05:38:33.9 & -02:36:26.3 & 19.16 &   &   \\
SigOri-SofI-13 & 05:38:34.0 & -02:36:36.9 & 13.68 & 05383405-0236375 & 165257 \\
SigOri-SofI-14 & 05:38:34.0 & -02:36:37.0 & 10.95 & 05383405-0236375 & 165257 \\
SigOri-SofI-15 & 05:38:34.2 & -02:34:49.8 & 19.52 &   &   \\
SigOri-SofI-16 & 05:38:34.6 & -02:38:26.1 & 16.88 &   &   \\
SigOri-SofI-17 & 05:38:34.8 & -02:34:27.3 & 17.35 &   &   \\
SigOri-SofI-18 & 05:38:34.8 & -02:36:20.0 & 16.59 & 05383491-0236206 &   \\
SigOri-SofI-19 & 05:38:34.9 & -02:36:20.1 & 14.12 & 05383491-0236206 &   \\
SigOri-SofI-20 & 05:38:35.0 & -02:34:55.5 & 17.97 & 05383510-0234559 &   \\
SigOri-SofI-21 & 05:38:35.1 & -02:34:55.9 & 16.48 & 05383510-0234559 &   \\
SigOri-SofI-22 & 05:38:35.1 & -02:35:57.8 & 18.25 &   &   \\
SigOri-SofI-23 & 05:38:35.2 & -02:34:09.2 & 19.54 &   &   \\
SigOri-SofI-24 & 05:38:35.2 & -02:34:21.4 & 18.72 &   &   \\
SigOri-SofI-25 & 05:38:35.2 & -02:34:36.8 & 17.69 &   &   \\
SigOri-SofI-26 & 05:38:35.2 & -02:34:37.4 & 16.58 &   &   \\
SigOri-SofI-27 & 05:38:35.2 & -02:34:51.9 & 18.09 &   &   \\
SigOri-SofI-28 & 05:38:35.3 & -02:36:02.1 & 17.01 &   &   \\
SigOri-SofI-29 & 05:38:35.3 & -02:38:37.1 & 17.42 & 05383540-0238373 &   \\
SigOri-SofI-30 & 05:38:35.4 & -02:35:47.1 & 16.80 &   &   \\
SigOri-SofI-31 & 05:38:35.4 & -02:38:37.0 & 14.95 & 05383540-0238373 &   \\
SigOri-SofI-32 & 05:38:35.5 & -02:35:34.4 & 18.93 &   &   \\
SigOri-SofI-33 & 05:38:35.6 & -02:34:26.1 & 18.32 &   &   \\
SigOri-SofI-34 & 05:38:35.7 & -02:34:26.8 & 17.07 &   &   \\
SigOri-SofI-35 & 05:38:35.9 & -02:35:01.7 & 19.08 &   &   \\
SigOri-SofI-36 & 05:38:36.0 & -02:38:08.9 & 16.08 &   &   \\
SigOri-SofI-37 & 05:38:36.1 & -02:34:27.0 & 18.91 &   &   \\
SigOri-SofI-38 & 05:38:36.3 & -02:36:28.2 & 18.45 &   &   \\
SigOri-SofI-39 & 05:38:36.4 & -02:37:32.1 & 16.22 &   &   \\
SigOri-SofI-40 & 05:38:36.5 & -02:33:19.7 & 18.81 &   &   \\
SigOri-SofI-41 & 05:38:36.5 & -02:35:33.4 & 18.75 &   &   \\
SigOri-SofI-42 & 05:38:36.7 & -02:33:22.7 & 19.14 &   &   \\
SigOri-SofI-43 & 05:38:36.7 & -02:34:57.1 & 18.81 &   &   \\
SigOri-SofI-44 & 05:38:36.8 & -02:35:16.0 & 12.70 & 05383686-0235163 &   \\
SigOri-SofI-45 & 05:38:36.8 & -02:36:42.8 & 12.00 & 05383687-0236432 & 126250 \\
SigOri-SofI-46 & 05:38:37.0 & -02:35:15.1 & 15.51 & 05383686-0235163 &   \\
SigOri-SofI-47 & 05:38:37.1 & -02:37:02.7 & 17.06 & 05383721-0237031 &   \\
SigOri-SofI-48 & 05:38:37.1 & -02:37:05.8 & 19.21 &   &   \\
SigOri-SofI-49 & 05:38:37.2 & -02:35:31.8 & 17.86 &   &   \\
SigOri-SofI-50 & 05:38:37.2 & -02:36:24.7 & 17.78 &   &   \\
SigOri-SofI-51 & 05:38:37.2 & -02:37:02.7 & 13.78 & 05383721-0237031 &   \\
SigOri-SofI-52 & 05:38:37.3 & -02:33:52.7 & 19.20 &   &   \\
SigOri-SofI-53 & 05:38:37.3 & -02:36:10.2 & 19.27 &   &   \\
SigOri-SofI-54 & 05:38:37.3 & -02:37:16.1 & 18.57 &   &   \\
SigOri-SofI-55 & 05:38:37.5 & -02:37:34.0 & 18.55 &   &   \\
SigOri-SofI-56 & 05:38:37.6 & -02:34:31.5 & 17.40 &   &   \\
SigOri-SofI-57 & 05:38:37.6 & -02:36:52.0 & 20.21 & 05383764-0236523 &   \\
SigOri-SofI-58 & 05:38:37.7 & -02:36:00.2 & 17.41 &   &   \\
SigOri-SofI-59 & 05:38:37.8 & -02:33:32.8 & 17.60 &   &   \\
SigOri-SofI-60 & 05:38:37.8 & -02:34:33.2 & 18.77 &   &   \\
SigOri-SofI-61 & 05:38:37.8 & -02:35:06.6 & 17.03 &   &   \\
SigOri-SofI-62 & 05:38:37.8 & -02:35:48.7 & 18.63 &   &   \\
SigOri-SofI-63 & 05:38:37.9 & -02:37:27.6 & 16.64 &   &   \\
SigOri-SofI-64 & 05:38:38.0 & -02:34:48.0 & 18.35 &   &   \\
SigOri-SofI-65 & 05:38:38.1 & -02:35:07.3 & 19.20 &   &   \\
SigOri-SofI-66 & 05:38:38.1 & -02:35:41.6 & 15.45 & 05383808-0235418 &   \\
SigOri-SofI-67 & 05:38:38.1 & -02:35:51.5 & 17.85 &   &   \\
SigOri-SofI-68 & 05:38:38.1 & -02:36:09.6 & 15.51 & 05383814-0236099 &   \\
SigOri-SofI-69 & 05:38:38.2 & -02:36:30.0 & 15.52 & 05383823-0236306 &   \\
SigOri-SofI-70 & 05:38:38.2 & -02:36:38.0 & 10.20 & 05383822-0236384 & 105249 \\
SigOri-SofI-71 & 05:38:38.2 & -02:38:24.5 & 18.84 &   &   \\
SigOri-SofI-72 & 05:38:38.2 & -02:38:43.8 & 16.53 &   &   \\
SigOri-SofI-73 & 05:38:38.4 & -02:34:45.9 & 15.78 & 05383845-0234466 &   \\
SigOri-SofI-74 & 05:38:38.4 & -02:36:31.4 & 19.22 & 05383823-0236306 &   \\
SigOri-SofI-75 & 05:38:38.4 & -02:38:05.3 & 15.78 & 05383840-0238054 &   \\
SigOri-SofI-76 & 05:38:38.5 & -02:34:36.2 & 16.05 & 05383858-0234371 &   \\
SigOri-SofI-77 & 05:38:38.5 & -02:34:54.6 & 9.64 & 05383848-0234550 & 114305 \\
SigOri-SofI-78 & 05:38:38.5 & -02:36:35.4 & 18.05 &   &   \\
SigOri-SofI-79 & 05:38:38.6 & -02:34:11.9 & 18.29 &   &   \\
SigOri-SofI-80 & 05:38:38.6 & -02:34:36.8 & 14.44 & 05383858-0234371 &   \\
SigOri-SofI-81 & 05:38:38.7 & -02:34:02.3 & 17.30 & 05383877-0234033 &   \\
SigOri-SofI-82 & 05:38:38.7 & -02:36:50.7 & 17.81 & 05383881-0236512 &   \\
SigOri-SofI-83 & 05:38:38.7 & -02:37:21.8 & 18.96 &   &   \\
SigOri-SofI-84 & 05:38:38.7 & -02:37:23.2 & 18.87 &   &   \\
SigOri-SofI-85 & 05:38:38.7 & -02:38:29.8 & 18.57 &   &   \\
SigOri-SofI-86 & 05:38:38.8 & -02:34:03.0 & 15.71 & 05383877-0234033 &   \\
SigOri-SofI-87 & 05:38:38.8 & -02:36:50.9 & 14.19 & 05383881-0236512 &   \\
SigOri-SofI-88 & 05:38:38.8 & -02:38:43.1 & 19.12 &   &   \\
SigOri-SofI-89 & 05:38:38.9 & -02:35:01.0 & 19.17 &   &   \\
SigOri-SofI-90 & 05:38:38.9 & -02:38:11.5 & 18.76 &   &   \\
SigOri-SofI-91 & 05:38:39.0 & -02:33:24.8 & 16.25 & 05383903-0233258 &   \\
SigOri-SofI-92 & 05:38:39.0 & -02:33:25.5 & 14.73 & 05383903-0233258 &   \\
SigOri-SofI-93 & 05:38:39.1 & -02:34:59.5 & 15.69 & 05383911-0235001 &   \\
SigOri-SofI-94 & 05:38:39.1 & -02:35:22.6 & 16.34 &   &   \\
SigOri-SofI-95 & 05:38:39.2 & -02:35:15.8 & 19.14 &   &   \\
SigOri-SofI-96 & 05:38:39.2 & -02:37:31.0 & 16.06 & 05383923-0237314 &   \\
SigOri-SofI-97 & 05:38:39.2 & -02:37:48.2 & 16.67 &   &   \\
SigOri-SofI-98 & 05:38:39.3 & -02:34:29.1 & 18.80 &   &   \\
SigOri-SofI-99 & 05:38:39.6 & -02:34:35.5 & 18.99 &   &   \\
SigOri-SofI-100 & 05:38:39.6 & -02:34:36.1 & 17.33 &   &   \\
SigOri-SofI-101 & 05:38:39.6 & -02:36:14.3 & 18.21 &   &   \\
SigOri-SofI-102 & 05:38:39.6 & -02:38:12.4 & 18.13 &   &   \\
SigOri-SofI-103 & 05:38:39.7 & -02:36:46.8 & 15.37 & 05383968-0236468 &   \\
SigOri-SofI-104 & 05:38:39.8 & -02:36:41.2 & 18.48 &   &   \\
SigOri-SofI-105 & 05:38:39.9 & -02:38:37.1 & 19.18 &   &   \\
SigOri-SofI-106 & 05:38:40.0 & -02:34:58.6 & 18.53 &   &   \\
SigOri-SofI-107 & 05:38:40.1 & -02:37:44.9 & 18.15 &   &   \\
SigOri-SofI-108 & 05:38:40.1 & -02:37:46.0 & 17.04 &   &   \\
SigOri-SofI-109 & 05:38:40.2 & -02:33:06.3 & 16.37 & 05384026-0233074 &   \\
SigOri-SofI-110 & 05:38:40.2 & -02:33:07.0 & 14.92 & 05384026-0233074 &   \\
SigOri-SofI-111 & 05:38:40.2 & -02:34:03.6 & 18.99 &   &   \\
SigOri-SofI-112 & 05:38:40.2 & -02:36:14.8 & 15.69 & 05384025-0236145 &   \\
SigOri-SofI-113 & 05:38:40.3 & -02:35:18.1 & 19.13 &   &   \\
SigOri-SofI-114 & 05:38:40.3 & -02:36:59.7 & 14.25 & 05384029-0237000 &   \\
SigOri-SofI-115 & 05:38:40.4 & -02:38:10.5 & 16.59 &   &   \\
SigOri-SofI-116 & 05:38:40.4 & -02:38:43.7 & 15.27 & 05384044-0238439 &   \\
SigOri-SofI-117 & 05:38:40.5 & -02:33:26.4 & 13.55 & 05384053-0233275 & 165337 \\
SigOri-SofI-118 & 05:38:40.5 & -02:33:27.1 & 12.09 & 05384053-0233275 & 165337 \\
SigOri-SofI-119 & 05:38:40.6 & -02:37:11.4 & 17.57 &   &   \\
SigOri-SofI-120 & 05:38:40.8 & -02:37:37.5 & 17.62 &   &   \\
SigOri-SofI-121 & 05:38:41.0 & -02:36:05.7 & 18.93 &   &   \\
SigOri-SofI-122 & 05:38:41.1 & -02:37:59.5 & 18.61 &   &   \\
SigOri-SofI-123 & 05:38:41.2 & -02:33:16.1 & 17.95 &   &   \\
SigOri-SofI-124 & 05:38:41.2 & -02:37:37.3 & 13.89 & 05384123-0237377 & 111208 \\
SigOri-SofI-125 & 05:38:41.2 & -02:38:10.2 & 16.74 &   &   \\
SigOri-SofI-126 & 05:38:41.3 & -02:35:53.3 & 16.85 & 05384146-0235523 & 50279 \\
SigOri-SofI-127 & 05:38:41.3 & -02:36:25.0 & 16.06 & 05384142-0236250 &   \\
SigOri-SofI-128 & 05:38:41.3 & -02:36:44.2 & 11.94 & 05384135-0236444 &   \\
SigOri-SofI-129 & 05:38:41.3 & -02:36:50.9 & 18.15 &   &   \\
SigOri-SofI-130 & 05:38:41.3 & -02:37:22.2 & 10.43 & 05384129-0237225 & 97212 \\
SigOri-SofI-131 & 05:38:41.3 & -02:38:32.5 & 18.16 &   &   \\
SigOri-SofI-132 & 05:38:41.4 & -02:33:42.2 & 17.86 &   &   \\
SigOri-SofI-133 & 05:38:41.4 & -02:35:52.0 & 12.85 & 05384146-0235523 & 50279 \\
SigOri-SofI-134 & 05:38:41.5 & -02:33:59.1 & 17.62 & 05384156-0234000 &   \\
SigOri-SofI-135 & 05:38:41.5 & -02:33:59.8 & 15.94 & 05384156-0234000 &   \\
SigOri-SofI-136 & 05:38:41.5 & -02:37:35.7 & 18.91 &   & 111208 \\
SigOri-SofI-137 & 05:38:41.5 & -02:37:41.5 & 19.34 &   &   \\
SigOri-SofI-138 & 05:38:41.6 & -02:34:40.8 & 17.31 &   &   \\
SigOri-SofI-139 & 05:38:41.6 & -02:36:50.2 & 18.77 &   &   \\
SigOri-SofI-140 & 05:38:41.6 & -02:38:43.8 & 17.80 &   &   \\
SigOri-SofI-141 & 05:38:41.7 & -02:33:12.1 & 17.20 &   &   \\
SigOri-SofI-142 & 05:38:41.8 & -02:34:28.9 & 17.18 &   & 111335  \\
SigOri-SofI-143 & 05:38:41.8 & -02:35:02.1 & 16.35 & 05384182-0235022 &   \\
SigOri-SofI-144 & 05:38:41.8 & -02:37:17.5 & 16.12 &   &   \\
SigOri-SofI-145 & 05:38:41.8 & -02:38:26.8 & 17.74 &   &   \\
SigOri-SofI-146 & 05:38:41.9 & -02:33:18.8 & 18.92 &   &   \\
SigOri-SofI-147 & 05:38:41.9 & -02:35:25.2 & 19.16 &   &   \\
SigOri-SofI-148 & 05:38:41.9 & -02:38:17.8 & 18.99 &   &   \\
SigOri-SofI-149 & 05:38:41.9 & -02:38:39.6 & 18.26 &   &   \\
SigOri-SofI-150 & 05:38:42.0 & -02:35:51.0 & 15.64 &   &   \\
SigOri-SofI-151 & 05:38:42.1 & -02:34:50.4 & 18.46 &   &   \\
SigOri-SofI-152 & 05:38:42.1 & -02:34:54.3 & 18.95 &   &   \\
SigOri-SofI-153 & 05:38:42.2 & -02:33:30.9 & 17.20 &   &   \\
SigOri-SofI-154 & 05:38:42.2 & -02:37:14.4 & 10.63 & 05384227-0237147 & 83207 \\
SigOri-SofI-155 & 05:38:42.2 & -02:38:02.8 & 18.64 &   &   \\
SigOri-SofI-156 & 05:38:42.2 & -02:38:53.3 & 17.79 &   &   \\
SigOri-SofI-157 & 05:38:42.3 & -02:33:34.7 & 16.81 &   &   \\
SigOri-SofI-158 & 05:38:42.3 & -02:36:14.6 & 17.56 &   &   \\
SigOri-SofI-159 & 05:38:42.3 & -02:38:10.0 & 17.45 &   &   \\
SigOri-SofI-160 & 05:38:42.4 & -02:34:10.1 & 17.44 &   &   \\
SigOri-SofI-161 & 05:38:42.4 & -02:35:27.5 & 19.26 &   &   \\
SigOri-SofI-162 & 05:38:42.4 & -02:36:04.1 & 13.09 & 05384239-0236044 & 36263 \\
SigOri-SofI-163 & 05:38:42.6 & -02:34:47.3 & 17.33 &   &   \\
SigOri-SofI-164 & 05:38:42.6 & -02:38:13.2 & 18.14 &   &   \\
SigOri-SofI-165 & 05:38:42.6 & -02:38:15.8 & 16.72 &   &   \\
SigOri-SofI-166 & 05:38:42.7 & -02:35:22.6 & 17.57 &   &   \\
SigOri-SofI-167 & 05:38:42.7 & -02:38:01.6 & 18.80 &   &   \\
SigOri-SofI-168 & 05:38:42.7 & -02:38:28.6 & 17.58 &   &   \\
SigOri-SofI-169 & 05:38:42.8 & -02:36:03.6 & 18.73 &   &   \\
SigOri-SofI-170 & 05:38:42.8 & -02:38:52.3 & 12.67 & 05384285-0238525 &   \\
SigOri-SofI-171 & 05:38:42.9 & -02:34:03.5 & 18.90 &   &   \\
SigOri-SofI-172 & 05:38:42.9 & -02:37:24.6 & 17.36 &   &   \\
SigOri-SofI-173 & 05:38:43.0 & -02:33:16.0 & 18.78 &   &   \\
SigOri-SofI-174 & 05:38:43.0 & -02:35:26.4 & 18.24 &   &   \\
SigOri-SofI-175 & 05:38:43.0 & -02:36:14.3 & 10.59 & 05384301-0236145 & 30241 \\
SigOri-SofI-176 & 05:38:43.0 & -02:36:26.9 & 18.42 &   &   \\
SigOri-SofI-177 & 05:38:43.2 & -02:37:22.9 & 18.07 &   &   \\
SigOri-SofI-178 & 05:38:43.3 & -02:33:17.1 & 16.78 &   &   \\
SigOri-SofI-179 & 05:38:43.3 & -02:36:47.7 & 16.19 &   &   \\
SigOri-SofI-180 & 05:38:43.4 & -02:34:49.8 & 18.84 &   &   \\
SigOri-SofI-181 & 05:38:43.4 & -02:34:50.6 & 17.61 &   & 72345  \\
SigOri-SofI-182 & 05:38:43.5 & -02:33:24.0 & 13.38 & 05384355-0233253 & 156353 \\
SigOri-SofI-183 & 05:38:43.5 & -02:33:24.9 & 10.73 & 05384355-0233253 & 156353 \\
SigOri-SofI-184 & 05:38:43.5 & -02:34:24.4 & 15.41 & 05384356-0234247 &   \\
SigOri-SofI-185 & 05:38:43.5 & -02:34:32.5 & 17.63 &   &   \\
SigOri-SofI-186 & 05:38:43.5 & -02:37:56.6 & 18.44 &   &   \\
SigOri-SofI-187 & 05:38:43.6 & -02:34:13.0 & 16.88 &   &   \\
SigOri-SofI-188 & 05:38:43.6 & -02:37:58.7 & 17.58 &   &   \\
SigOri-SofI-189 & 05:38:43.6 & -02:38:33.0 & 19.01 &   &   \\
SigOri-SofI-190 & 05:38:43.8 & -02:34:42.3 & 17.68 &   &   \\
SigOri-SofI-191 & 05:38:43.8 & -02:37:06.5 & 11.53 & 05384386-0237068 & 68191 \\
SigOri-SofI-192 & 05:38:43.9 & -02:33:42.3 & 17.89 &   &   \\
SigOri-SofI-193 & 05:38:43.9 & -02:34:08.6 & 18.90 &   &   \\
SigOri-SofI-194 & 05:38:43.9 & -02:36:39.6 & 16.97 &   &   \\
SigOri-SofI-195 & 05:38:44.1 & -02:37:10.2 & 19.53 &   & 68191 \\
SigOri-SofI-196 & 05:38:44.2 & -02:34:55.3 & 18.08 &   &   \\
SigOri-SofI-197 & 05:38:44.3 & -02:38:14.4 & 18.83 &   &   \\
SigOri-SofI-198 & 05:38:44.4 & -02:34:12.4 & 18.28 &   &   \\
SigOri-SofI-199 & 05:38:44.4 & -02:37:35.8 & 18.50 &   &   \\
SigOri-SofI-200 & 05:38:44.4 & -02:38:20.1 & 18.61 &   &   \\
SigOri-SofI-201 & 05:38:44.5 & -02:33:33.8 & 18.61 &   &   \\
SigOri-SofI-202 & 05:38:44.6 & -02:33:17.6 & 18.18 &   &   \\
SigOri-SofI-203 & 05:38:44.7 & -02:35:15.2 & 16.89 &   &   \\
SigOri-SofI-204 & 05:38:44.8 & -02:33:57.2 & 9.64 & 05384480-0233576 & 123000 \\
SigOri-SofI-205 & 05:38:44.8 & -02:36:56.9 & 16.50 &   &   \\
SigOri-SofI-206 & 05:38:44.9 & -02:36:41.2 & 17.03 &   &   \\
SigOri-SofI-207 & 05:38:45.0 & -02:33:23.9 & 14.80 &   &   \\
SigOri-SofI-208 & 05:38:45.0 & -02:33:38.8 & 18.08 &   &   \\
SigOri-SofI-209 & 05:38:45.0 & -02:36:32.6 & 18.68 &   &   \\
SigOri-SofI-210 & 05:38:45.1 & -02:34:10.8 & 18.24 &   &   \\
SigOri-SofI-211 & 05:38:45.1 & -02:36:41.4 & 17.81 &   &   \\
SigOri-SofI-212 & 05:38:45.1 & -02:38:00.0 & 17.76 &   &   \\
SigOri-SofI-213 & 05:38:45.1 & -02:38:18.1 & 15.49 & 05384516-0238181 &   \\
SigOri-SofI-214 & 05:38:45.1 & -02:38:36.5 & 17.86 &   &   \\
SigOri-SofI-215 & 05:38:45.1 & -02:38:41.4 & 18.34 &   &   \\
SigOri-SofI-216 & 05:38:45.2 & -02:37:29.0 & 12.02 & 05384527-0237292 & 89175 \\
SigOri-SofI-217 & 05:38:45.3 & -02:37:32.5 & 19.37 &   & 89175 \\
SigOri-SofI-218 & 05:38:45.5 & -02:34:07.1 & 18.88 &   &   \\
SigOri-SofI-219 & 05:38:45.7 & -02:33:58.1 & 18.01 &   &   \\
SigOri-SofI-220 & 05:38:45.7 & -02:34:10.6 & 18.20 &   &   \\
SigOri-SofI-221 & 05:38:45.7 & -02:34:11.3 & 17.92 &   &   \\
SigOri-SofI-222 & 05:38:45.7 & -02:37:32.5 & 12.45 & 05384571-0237327 &   \\
SigOri-SofI-223 & 05:38:45.7 & -02:38:26.4 & 17.67 &   &   \\
SigOri-SofI-224 & 05:38:45.8 & -02:37:13.8 & 19.24 &   &   \\
SigOri-SofI-225 & 05:38:45.8 & -02:38:50.2 & 18.24 &   &   \\
SigOri-SofI-226 & 05:38:45.8 & -02:38:51.8 & 17.23 &   &   \\
SigOri-SofI-227 & 05:38:45.9 & -02:36:39.4 & 19.11 &   &   \\
SigOri-SofI-228 & 05:38:46.0 & -02:37:48.6 & 17.66 &   &   \\
SigOri-SofI-229 & 05:38:46.1 & -02:34:16.8 & 19.15 &   &   \\
SigOri-SofI-230 & 05:38:46.2 & -02:34:14.7 & 16.68 &   &   \\
SigOri-SofI-231 & 05:38:46.2 & -02:35:16.3 & 17.63 &   &   \\
SigOri-SofI-232 & 05:38:46.2 & -02:37:10.0 & 14.01 & 05384626-0237102 &   \\
SigOri-SofI-233 & 05:38:46.3 & -02:34:22.1 & 16.29 &   &   \\
SigOri-SofI-234 & 05:38:46.3 & -02:34:53.4 & 17.32 &   &   \\
SigOri-SofI-235 & 05:38:46.3 & -02:37:13.3 & 18.89 &   &   \\
SigOri-SofI-236 & 05:38:46.4 & -02:34:33.4 & 14.52 & 05384639-0234336 &   \\
SigOri-SofI-237 & 05:38:46.6 & -02:34:58.1 & 16.24 &   &   \\
SigOri-SofI-238 & 05:38:46.8 & -02:36:43.3 & 12.21 & 05384684-0236435 & 53144 \\
SigOri-SofI-239 & 05:38:46.9 & -02:38:09.8 & 17.74 &   &   \\
SigOri-SofI-240 & 05:38:47.0 & -02:36:27.3 & 19.29 &   &   \\
SigOri-SofI-241 & 05:38:47.0 & -02:37:01.2 & 17.30 &   &   \\
SigOri-SofI-242 & 05:38:47.0 & -02:37:56.5 & 18.04 &   &   \\
SigOri-SofI-243 & 05:38:47.1 & -02:33:15.1 & 19.26 &   &   \\
SigOri-SofI-244 & 05:38:47.1 & -02:36:28.1 & 18.75 &   &   \\
SigOri-SofI-245 & 05:38:47.1 & -02:38:29.6 & 18.68 &   &   \\
SigOri-SofI-246 & 05:38:47.1 & -02:38:36.7 & 19.27 &   &   \\
SigOri-SofI-247 & 05:38:47.1 & -02:39:02.1 & 17.13 &   &   \\
SigOri-SofI-248 & 05:38:47.2 & -02:34:36.5 & 11.29 & 05384718-0234368 & 91024 \\
SigOri-SofI-249 & 05:38:47.2 & -02:37:34.2 & 16.36 &   &   \\
SigOri-SofI-250 & 05:38:47.2 & -02:37:38.1 & 19.37 &   &   \\
SigOri-SofI-251 & 05:38:47.2 & -02:39:02.6 & 16.38 &   &   \\
SigOri-SofI-252 & 05:38:47.2 & -02:39:06.4 & 18.18 &   &   \\
SigOri-SofI-253 & 05:38:47.3 & -02:34:03.2 & 19.04 &   &   \\
SigOri-SofI-254 & 05:38:47.3 & -02:35:19.0 & 15.43 &   &   \\
SigOri-SofI-255 & 05:38:47.3 & -02:38:42.9 & 18.88 &   &   \\
SigOri-SofI-256 & 05:38:47.5 & -02:33:59.3 & 19.18 &   &   \\
SigOri-SofI-257 & 05:38:47.5 & -02:35:18.8 & 14.78 &   &   \\
SigOri-SofI-258 & 05:38:47.5 & -02:38:24.2 & 16.64 &   &   \\
SigOri-SofI-259 & 05:38:47.6 & -02:34:17.6 & 18.88 &   &   \\
SigOri-SofI-260 & 05:38:47.6 & -02:38:00.0 & 19.02 &   &   \\
SigOri-SofI-261 & 05:38:47.7 & -02:36:40.5 & 15.94 & 05384771-0236407 &   \\
SigOri-SofI-262 & 05:38:47.7 & -02:36:41.8 & 19.07 & 05384771-0236407 &   \\
SigOri-SofI-263 & 05:38:47.8 & -02:34:52.9 & 19.40 &   &   \\
SigOri-SofI-264 & 05:38:47.8 & -02:36:43.7 & 18.78 &   &   \\
SigOri-SofI-265 & 05:38:47.9 & -02:37:19.2 & 10.76 & 05384791-0237192 & 92149 \\
SigOri-SofI-266 & 05:38:48.0 & -02:34:00.8 & 18.33 &   &   \\
SigOri-SofI-267 & 05:38:48.0 & -02:37:18.2 & 11.58 & 05384791-0237192 & 92149 \\
SigOri-SofI-268 & 05:38:48.1 & -02:33:08.4 & 17.49 &   &   \\
SigOri-SofI-269 & 05:38:48.1 & -02:34:15.9 & 17.04 &   &   \\
SigOri-SofI-270 & 05:38:48.1 & -02:36:28.3 & 16.88 &   &   \\
SigOri-SofI-271 & 05:38:48.1 & -02:38:42.6 & 17.78 &   &   \\
SigOri-SofI-272 & 05:38:48.3 & -02:33:47.1 & 17.68 &   &   \\
SigOri-SofI-273 & 05:38:48.3 & -02:35:48.3 & 19.52 &   &   \\
SigOri-SofI-274 & 05:38:48.3 & -02:36:40.8 & 10.99 & 05384828-0236409 & 67128 \\
SigOri-SofI-275 & 05:38:48.3 & -02:37:04.8 & 19.29 &   &   \\
SigOri-SofI-276 & 05:38:48.3 & -02:37:15.8 & 18.50 &   &   \\
SigOri-SofI-277 & 05:38:48.3 & -02:37:22.5 & 19.37 &   &   \\
SigOri-SofI-278 & 05:38:48.3 & -02:38:42.7 & 19.31 &   &   \\
SigOri-SofI-279 & 05:38:48.4 & -02:33:46.3 & 18.84 &   &   \\
SigOri-SofI-280 & 05:38:48.5 & -02:33:55.4 & 17.14 &   &   \\
SigOri-SofI-281 & 05:38:48.5 & -02:36:07.7 & 18.35 &   &   \\
SigOri-SofI-282 & 05:38:48.5 & -02:37:02.1 & 17.91 &   &   \\
SigOri-SofI-283 & 05:38:48.5 & -02:38:56.2 & 18.81 &   &   \\
SigOri-SofI-284 & 05:38:48.6 & -02:33:37.0 & 18.05 &   &   \\
SigOri-SofI-285 & 05:38:48.6 & -02:36:51.9 & 17.67 &   &   \\
SigOri-SofI-286 & 05:38:48.7 & -02:36:01.9 & 19.01 &   &   \\
SigOri-SofI-287 & 05:38:48.7 & -02:36:16.0 & 11.13 & 05384868-0236162 & 61105 \\
SigOri-SofI-288 & 05:38:48.7 & -02:36:27.1 & 19.05 &   &   \\
SigOri-SofI-289 & 05:38:48.7 & -02:36:46.0 & 18.70 &   &   \\
SigOri-SofI-290 & 05:38:48.8 & -02:36:42.2 & 18.32 &   &   \\
SigOri-SofI-291 & 05:38:48.8 & -02:38:56.5 & 17.71 &   &   \\
SigOri-SofI-292 & 05:38:48.9 & -02:33:28.6 & 18.59 &   &   \\
SigOri-SofI-293 & 05:38:48.9 & -02:33:56.9 & 19.29 &   &   \\
SigOri-SofI-294 & 05:38:49.0 & -02:38:36.2 & 14.27 & 05384906-0238364 &   \\
SigOri-SofI-295 & 05:38:49.0 & -02:38:42.1 & 17.39 &   &   \\
SigOri-SofI-296 & 05:38:49.1 & -02:38:22.1 & 10.70 & 05384917-0238222 & 157155 \\
SigOri-SofI-297 & 05:38:49.2 & -02:33:49.9 & 19.11 &   &   \\
SigOri-SofI-298 & 05:38:49.2 & -02:35:49.4 & 17.05 &   &   \\
SigOri-SofI-299 & 05:38:49.4 & -02:38:26.9 & 19.54 &   &   \\
SigOri-SofI-300 & 05:38:49.6 & -02:34:55.7 & 17.69 &   & 100048 \\
SigOri-SofI-301 & 05:38:49.6 & -02:37:17.8 & 18.70 &   &   \\
SigOri-SofI-302 & 05:38:49.7 & -02:34:52.3 & 12.12 & 05384970-0234526 & 100048 \\
SigOri-SofI-303 & 05:38:49.7 & -02:36:07.1 & 18.82 &   &   \\
SigOri-SofI-304 & 05:38:49.7 & -02:38:56.8 & 16.74 &   &   \\
SigOri-SofI-305 & 05:38:50.0 & -02:34:35.9 & 14.90 & 05385001-0234361 &   \\
SigOri-SofI-306 & 05:38:50.0 & -02:37:35.3 & 11.94 & 05385003-0237354 & 124140 \\
SigOri-SofI-307 & 05:38:50.1 & -02:34:43.2 & 15.12 & 05385008-0234433 &   \\
SigOri-SofI-308 & 05:38:50.2 & -02:36:51.2 & 18.78 &   &   \\
SigOri-SofI-309 & 05:38:50.2 & -02:37:00.6 & 18.64 &   &   \\
SigOri-SofI-310 & 05:38:50.3 & -02:36:21.2 & 18.59 &   &   \\
SigOri-SofI-311 & 05:38:50.3 & -02:36:46.7 & 18.54 &   &   \\
SigOri-SofI-312 & 05:38:50.3 & -02:37:03.7 & 19.02 &   &   \\
SigOri-SofI-313 & 05:38:50.3 & -02:38:59.4 & 17.42 &   &   \\
SigOri-SofI-314 & 05:38:50.4 & -02:34:09.3 & 18.04 &   &   \\
SigOri-SofI-315 & 05:38:50.4 & -02:34:28.9 & 18.10 &   &   \\
SigOri-SofI-316 & 05:38:50.4 & -02:35:09.1 & 13.86 & 05385043-0235093 &   \\
SigOri-SofI-317 & 05:38:50.4 & -02:36:43.5 & 15.80 & 05385042-0236431 &   \\
SigOri-SofI-318 & 05:38:50.5 & -02:34:33.0 & 18.97 &   &   \\
SigOri-SofI-319 & 05:38:50.6 & -02:34:13.3 & 15.62 & 05385065-0234135 &   \\
SigOri-SofI-320 & 05:38:50.7 & -02:38:28.8 & 16.15 & 05385081-0238286 &   \\
SigOri-SofI-321 & 05:38:50.8 & -02:34:09.3 & 17.88 &   &   \\
SigOri-SofI-322 & 05:38:50.8 & -02:36:26.5 & 12.11 & 05385077-0236267 & 94106 \\
SigOri-SofI-323 & 05:38:50.9 & -02:34:00.3 & 17.04 &   &   \\
SigOri-SofI-324 & 05:38:50.9 & -02:36:39.6 & 16.97 &   &   \\
SigOri-SofI-325 & 05:38:51.0 & -02:34:01.6 & 19.00 &   &   \\
SigOri-SofI-326 & 05:38:51.1 & -02:33:56.0 & 18.48 &   &   \\
SigOri-SofI-327 & 05:38:51.1 & -02:36:22.4 & 19.13 &   &   \\
SigOri-SofI-328 & 05:38:51.2 & -02:33:55.8 & 17.41 &   &   \\
SigOri-SofI-329 & 05:38:51.2 & -02:38:55.9 & 17.02 &   &   \\
SigOri-SofI-330 & 05:38:51.4 & -02:34:38.4 & 19.17 &   &   \\
SigOri-SofI-331 & 05:38:51.4 & -02:36:20.4 & 11.46 & 05385145-0236205 & 102101 \\
SigOri-SofI-332 & 05:38:51.6 & -02:38:12.8 & 18.83 &   &   \\
SigOri-SofI-333 & 05:38:51.6 & -02:38:23.4 & 16.45 &   &   \\
SigOri-SofI-334 & 05:38:51.7 & -02:35:01.4 & 18.85 &   &   \\
SigOri-SofI-335 & 05:38:51.7 & -02:36:03.1 & 11.99 & 05385173-0236033 & 105092 \\
SigOri-SofI-336 & 05:38:51.7 & -02:38:10.0 & 18.03 &   &   \\
SigOri-SofI-337 & 05:38:51.8 & -02:36:36.4 & 18.64 &   &   \\
SigOri-SofI-338 & 05:38:51.8 & -02:37:25.9 & 19.19 &   &   \\
SigOri-SofI-339 & 05:38:51.8 & -02:38:26.9 & 16.05 &   &   \\
SigOri-SofI-340 & 05:38:51.9 & -02:33:23.9 & 17.23 &   &   \\
SigOri-SofI-341 & 05:38:51.9 & -02:34:13.2 & 15.00 & 05385186-0234134 &   \\
SigOri-SofI-342 & 05:38:52.1 & -02:33:23.3 & 19.05 &   &   \\
SigOri-SofI-343 & 05:38:52.1 & -02:35:39.9 & 18.06 & 05385212-0235402 &   \\
SigOri-SofI-344 & 05:38:52.2 & -02:37:15.7 & 18.87 &   &   \\
SigOri-SofI-345 & 05:38:52.3 & -02:39:07.7 & 18.27 &   &   \\
SigOri-SofI-346 & 05:38:52.4 & -02:37:16.8 & 19.45 &   &   \\
SigOri-SofI-347 & 05:38:52.5 & -02:37:37.7 & 17.99 &   &   \\
SigOri-SofI-348 & 05:38:52.6 & -02:35:42.5 & 18.34 &   &   \\
SigOri-SofI-349 & 05:38:52.6 & -02:36:50.0 & 18.84 &   &   \\
SigOri-SofI-350 & 05:38:52.6 & -02:37:12.3 & 18.31 &   &   \\
SigOri-SofI-351 & 05:38:52.7 & -02:35:51.8 & 18.92 &   &   \\
SigOri-SofI-352 & 05:38:52.8 & -02:35:33.2 & 17.47 &   &   \\
SigOri-SofI-353 & 05:38:52.8 & -02:36:39.1 & 17.79 &   &   \\
SigOri-SofI-354 & 05:38:52.8 & -02:37:50.4 & 18.51 &   &   \\
SigOri-SofI-355 & 05:38:52.8 & -02:38:30.5 & 17.14 &   &   \\
SigOri-SofI-356 & 05:38:53.0 & -02:34:42.4 & 18.83 &   &   \\
SigOri-SofI-357 & 05:38:53.0 & -02:37:24.2 & 19.39 &   &   \\
SigOri-SofI-358 & 05:38:53.1 & -02:35:14.4 & 18.04 &   &   \\
SigOri-SofI-359 & 05:38:53.1 & -02:37:58.7 & 18.03 &   &   \\
SigOri-SofI-360 & 05:38:53.2 & -02:38:08.1 & 19.04 &   &   \\
SigOri-SofI-361 & 05:38:53.2 & -02:38:47.2 & 18.54 &   &   \\
SigOri-SofI-362 & 05:38:53.3 & -02:34:32.2 & 17.03 &   &   \\
SigOri-SofI-363 & 05:38:53.3 & -02:35:18.0 & 19.09 &   &   \\
SigOri-SofI-364 & 05:38:53.4 & -02:33:22.6 & 9.98 & 05385337-0233229 & 203039 \\
SigOri-SofI-365 & 05:38:53.4 & -02:35:22.9 & 15.79 & 05385340-0235231 &   \\
SigOri-SofI-366 & 05:38:53.4 & -02:36:57.8 & 19.34 &   &   \\
SigOri-SofI-367 & 05:38:53.4 & -02:37:02.4 & 19.13 &   &   \\
SigOri-SofI-368 & 05:38:53.5 & -02:37:05.8 & 18.76 &   &   \\
SigOri-SofI-369 & 05:38:53.6 & -02:35:50.3 & 17.62 &   &   \\
SigOri-SofI-370 & 05:38:53.8 & -02:35:50.3 & 17.32 &   &   \\
SigOri-SofI-371 & 05:38:53.8 & -02:37:22.2 & 18.17 &   &   \\
SigOri-SofI-372 & 05:38:53.9 & -02:34:19.6 & 12.73 & 05385387-0234199 &   \\
SigOri-SofI-373 & 05:38:54.0 & -02:33:35.7 & 18.95 &   &   \\
SigOri-SofI-374 & 05:38:54.0 & -02:36:06.4 & 18.96 &   &   \\
SigOri-SofI-375 & 05:38:54.4 & -02:36:48.2 & 18.43 &   &   \\
SigOri-SofI-376 & 05:38:54.7 & -02:34:18.7 & 18.87 &   &   \\
SigOri-SofI-377 & 05:38:54.7 & -02:34:27.4 & 17.95 & 05385467-0234280 &   \\
SigOri-SofI-378 & 05:38:54.7 & -02:35:53.8 & 17.92 &   &   \\
SigOri-SofI-379 & 05:38:54.7 & -02:37:42.6 & 17.20 & 05385486-0237411 &   \\
SigOri-SofI-380 & 05:38:54.8 & -02:35:59.4 & 18.26 &   &   \\
SigOri-SofI-381 & 05:38:54.8 & -02:37:07.7 & 18.50 &   &   \\
SigOri-SofI-382 & 05:38:54.8 & -02:37:40.8 & 14.93 & 05385486-0237411 &   \\
SigOri-SofI-383 & 05:38:54.9 & -02:35:08.7 & 16.73 &   &   \\
SigOri-SofI-384 & 05:38:55.0 & -02:38:25.5 & 18.39 &   &   \\
SigOri-SofI-385 & 05:38:55.1 & -02:34:58.6 & 18.26 &   &   \\
SigOri-SofI-386 & 05:38:55.2 & -02:36:09.1 & 18.90 &   &   \\
SigOri-SofI-387 & 05:38:55.3 & -02:36:50.9 & 17.96 &   &   \\
SigOri-SofI-388 & 05:38:55.4 & -02:34:46.5 & 19.10 &   &   \\
SigOri-SofI-389 & 05:38:55.4 & -02:37:01.1 & 14.47 & 05385546-0237012 &   \\
SigOri-SofI-390 & 05:38:55.5 & -02:36:28.7 & 19.34 &   &   \\
SigOri-SofI-391 & 05:38:55.6 & -02:37:37.9 & 17.48 &   &   \\
SigOri-SofI-392 & 05:38:55.8 & -02:35:26.9 & 19.08 &   &   \\
SigOri-SofI-393 & 05:38:56.0 & -02:34:09.2 & 18.66 &   &   \\
SigOri-SofI-394 & 05:38:56.1 & -02:34:11.1 & 17.78 &   &   \\
SigOri-SofI-395 & 05:38:56.1 & -02:34:27.5 & 16.14 & 05385606-0234278 &   \\
SigOri-SofI-396 & 05:38:56.2 & -02:35:49.6 & 16.72 &   &   \\
SigOri-SofI-397 & 05:38:56.3 & -02:34:40.4 & 14.28 & 05385636-0234401 &   \\
SigOri-SofI-398 & 05:38:56.4 & -02:34:38.5 & 14.60 & 05385636-0234401 &   \\
\hline
\end{tabular}

Notes - The error on the photometry is dominated by the zeropoint uncertainty (0.19~mag). The 2MASS \citep{2006AJ....131.1163S} and Mayrit \citep{2008A&A...478..667C} nearest matches within 5\arcsec are indicated. Because of the higher spatial resolution of the SofI image compared to these two catalogs, the same 2MASS or Mayrit source is sometimes associated to several SofI sources.
\end{table*}
\end{center}

\end{document}